\documentclass[prc,showpacs,superscriptaddress,twocolumn]{revtex4}

\usepackage{graphicx}
\usepackage{amsmath}
\usepackage{amsfonts}
\usepackage{array}[=2016-10-06]

\begin{document}
	
	\title{Deep learning approaches to extract nuclear deformation parameters from initial-state information in heavy-ion collisions}
	
	\author{Jun-Qi Tao}
	\affiliation{Key Laboratory of Quark \& Lepton Physics (MOE) and Institute of Particle Physics, Central China Normal University, Wuhan 430079, China}
    \affiliation{School of Science and Engineering, The Chinese University of Hong Kong, Shenzhen (CUHK-Shenzhen), Guangdong, 518172, China}
	\affiliation{Helmholtz Research Academy Hesse for FAIR (HFHF), GSI Helmholtz Center for heavy-ion Research, Frankfurt am Main 60438, Germany}
	\affiliation{Institute for Theoretical Physics, Johann Wolfgang Goethe University, Frankfurt am Main 60438, Germany}

	\author{Yang Liu}
	\affiliation{Key Laboratory of Quark \& Lepton Physics (MOE) and Institute of Particle Physics, Central China Normal University, Wuhan 430079, China}
    
	\author{Yu Sha}
	\affiliation{School of Science and Engineering, The Chinese University of Hong Kong, Shenzhen (CUHK-Shenzhen), Guangdong, 518172, China}
    
	\author{Xiang Fan}
	\affiliation{Key Laboratory of Quark \& Lepton Physics (MOE) and Institute of Particle Physics, Central China Normal University, Wuhan 430079, China}
    
    \author{Yan-Sheng Tu}
    \affiliation{School of Science and Engineering, The Chinese University of Hong Kong, Shenzhen (CUHK-Shenzhen), Guangdong, 518172, China}
	
	\author{Kai Zhou}
	\email[]{zhoukai@cuhk.edu.cn}
	\affiliation{School of Science and Engineering, The Chinese University of Hong Kong, Shenzhen (CUHK-Shenzhen), Guangdong, 518172, China}
	\affiliation{School of Artificial Intelligence, The Chinese University of Hong Kong, Shenzhen (CUHK-Shenzhen), Guangdong, 518172, China}
    
	\author{Hua Zheng}
	\email[]{zhengh@snnu.edu.cn}
	\affiliation{School of Physics and Information Technology, Shaanxi Normal University, Xi’an 710119, China}
	
	\author{Ben-Wei Zhang}
	\email[]{bwzhang@mail.ccnu.edu.cn}
	\affiliation{Key Laboratory of Quark \& Lepton Physics (MOE) and Institute of Particle Physics, Central China Normal University, Wuhan 430079, China}

	\date{\today}
	
	\begin{abstract}
		The deformation of heavy nuclei leaves characteristic imprints on the initial conditions of relativistic heavy-ion collisions. However, event-by-event fluctuations make the quantitative extraction of this information challenging. This study examines the identifiability of the quadrupole ($\beta_2$) and hexadecapole ($\beta_4$) deformation parameters from nucleon configurations sampled from a deformed Woods–Saxon distribution commonly used in initial-state modeling of heavy-ion collisions. As a baseline, we first establish an upper bound on the ``intrinsic identifiability" of deformation information at the most microscopic level by constructing permutation-invariant point-cloud networks under controlled multi-event grouping. We then extend the analysis to the more realistic initial entropy-density profiles generated by the T\raisebox{-0.5ex}{R}ENTo model, where both standard regression and simulation-based inference (SBI) with conditional normalizing flows are employed to reconstruct the deformation parameters from ensembles of event images supplemented with global attributes. Multi-event averaging is found to be essential in this setting for suppressing stochastic fluctuations and revealing the underlying deformation information. While standard regression efficiently captures the central trends of deformation through point estimates, SBI provides calibrated posterior distributions, offering a more complete and robust characterization of uncertainty. Collectively, our results demonstrate that deformation information is effectively encoded in the initial-state and becomes increasingly identifiable with sufficient ensemble averaging, laying a solid foundation for future extensions toward more complete dynamical modeling and final-state observables.
	\end{abstract}
	
	
	\maketitle
	
	\section{Introduction}\label{Sec1}
	
	High-energy heavy-ion collisions provide a unique opportunity to study strongly interacting matter under extreme conditions. At the Relativistic Heavy Ion Collider (RHIC) and Large Hadron Collider (LHC), such collisions can reach temperatures and energy densities high enough to create the quark–gluon plasma (QGP), a deconfined state of quarks and gluons predicted by quantum chromodynamics (QCD) \cite{Braun-Munzinger2007,Shuryak2017,Adams2005}. The QGP exhibits strong collective flow and behaves like a nearly perfect fluid, offering a controlled environment in which the transport properties of hot and dense QCD matter can be investigated \cite{Romatschke2007,Ollitrault1992}. Meanwhile, the geometry and intrinsic structure of the colliding nuclei play a crucial role in determining the initial conditions of the collision dynamics \cite{Miller2007}. In particular, nuclear deformation parameters such as quadrupole ($\beta_2$) and hexadecapole ($\beta_4$) moments can imprint characteristic patterns on the initial entropy- and energy-density profile, which subsequently influences final-state observables including particle multiplicities and anisotropic flow \cite{Abdulhamid2024,Jia2022,Giacalone2021,Shou2015,Dimri2023}.
	
	Recently, we have witnessed a growing recognition that heavy-ion collisions can also serve as a novel probe of nuclear structure. Comparative studies of isobar systems such as \({}^{96}\mathrm{Ru}+{}^{96}\mathrm{Ru}\) and \({}^{96}\mathrm{Zr}+{}^{96}\mathrm{Zr}\) have revealed sensitivity to neutron-skin thickness and quadrupole deformation through flow coefficients and mean transverse momentum fluctuations \cite{nXu2022,Giacalone2021}. Scaling relations between deformation parameters and flow correlators have further opened a quantitative pathway toward imaging nuclear shapes through ultrarelativistic collisions \cite{nJia2022}. In parallel, Bayesian-imaging approaches have demonstrated that single-system observables already contain sufficient information to constrain Woods--Saxon parameters, while isobar ratios alone may lead to degeneracies unless complemented by multiplicity distributions \cite{Cheng2023}. More broadly, a recent paper by Jia \emph{et al.} \cite{nJia2024} outlines a program to use collider observables to constrain nuclear deformation, triaxiality and radial structure across the nuclide chart. Complementary studies of Xe-isotope collisions at the LHC have demonstrated sensitivity to \(\gamma\)-soft and triaxial deformation modes, for example in \({}^{129}\mathrm{Xe}+{}^{129}\mathrm{Xe}\) collisions \cite{nZhao2024}. Together, these developments establish relativistic heavy-ion collisions as a powerful cross-disciplinary bridge connecting traditional low-energy nuclear structure with high-energy QCD dynamics. 
    
    Inspired by these advances, the present work explores whether machine learning methods can further strengthen this connection by inferring the deformation parameters $(\beta_2,\beta_4)$ directly from initial-state entropy-density profiles. However, to rigorously interpret these structure-driven signatures and establish a robust link to experimentally accessible observables, realistic theoretical modeling of the collision process is indispensable. Over the past decades, a broad suite of theoretical and phenomenological tools has been developed to describe the full dynamical evolution of heavy-ion collisions. This includes initial-state generators (e.g., T\raisebox{-0.5ex}{R}ENTo \cite{Moreland2015}, IP-Glasma \cite{Schenke2012}), relativistic viscous hydrodynamics (MUSIC \cite{Schenke2010,Schenke2011,Paquet2016}, CLVisc \cite{Pang2018a,Jiang2022,Jiang2023,Jiang2025}) and hadronic transport models (UrQMD \cite{Bass1998,Bleicher1999}, SMASH \cite{Weil2016,Petersen2019,Weil2016a}). These packages provide essential baselines for understanding how the initial geometric information encoded in the initial-state propagates through the system's evolution to influence final-state observables which makes this work valuable. 
	
	Parallel to these theoretical developments, machine-learning (ML) techniques have emerged as a powerful complementary toolset in high-energy nuclear physics \cite{Carleo2019,Radovic2018,Du2020,Kuttan2021,Zhao2022,OmanaKuttan2023,Zhou2024,He2023,Oliveira2017,Steinheimer2019,Huang2022,Paganini2018,OmanaKuttan2020,OmanaKuttan:2024mwr,Sun2025,Pang2019,Pang2018,Tao2025}. ML-based methods excel at disentangling complex, non-linear correlations within high-dimensional data and in providing fast surrogates for computationally expensive simulations. Applications include event-by-event emulation \cite{Kuttan2025}, hydrodynamic surrogates \cite{Sun2025}, nuclear-structure inference in collision systems \cite{Pang2019} and QCD equation-of-state reconstruction \cite{Pang2018}. These advances underscore the growing role of machine-learning not merely as a technical aid, but as an integral component of modern physics research.
	
	In this study, we investigate the capability of machine-learning to infer nuclear deformation parameters $(\beta_2, \beta_4)$ directly from initial-state information for heavy-ion collisions. We begin by establishing an idealized baseline using nucleon configurations to verify the intrinsic identifiability at the microscopic level. Then we extend the analysis to realistic entropy-density profiles simulated from the T\raisebox{-0.5ex}{R}ENTo model. Although large fluctuations in nucleon sampling and entropy deposition pose a significant challenge to direct regression of the nuclear deformation, we demonstrate that simulation-based inference (SBI) \cite{Cranmer2020,Lueckmann2021} offers a promising avenue by reconstructing the full posterior distributions over the nuclear deformation parameters. This study provides an approximate upper bound on the extractable information about $(\beta_2, \beta_4)$ from heavy-ion collisions within this modeling framework, since subsequent hydrodynamic evolution inevitably blurs detailed geometric features due to viscous dissipation. Successfully identifying deformation in the initial-state serves as a necessary prerequisite for any experimental extraction from final-state observables.
	
	The paper is organized as follows. In Section II, we introduce the neural network architectures and training strategies used in two scenarios: nucleon configurations and initial entropy-density profiles. In Section \ref{Sec3}, we present the main results, including those of the baseline study of deformation inference from deformed Woods–Saxon nucleon configurations and the subsequent analysis based on initial entropy density profiles simulated with the T\raisebox{-0.5ex}{R}ENTo model. The performance of regression and SBI is compared and the implications for extracting nuclear deformation information from initial conditions are discussed. Finally, in Section \ref{Sec4}, we provide a brief conclusion and outlook.
	
	\section{Model framework}\label{Sec2}
	
	\subsection{Neural network for nucleon configuration(s)}
	
	\begin{figure*}[htbp]
		\centering
		\includegraphics[scale=0.6]{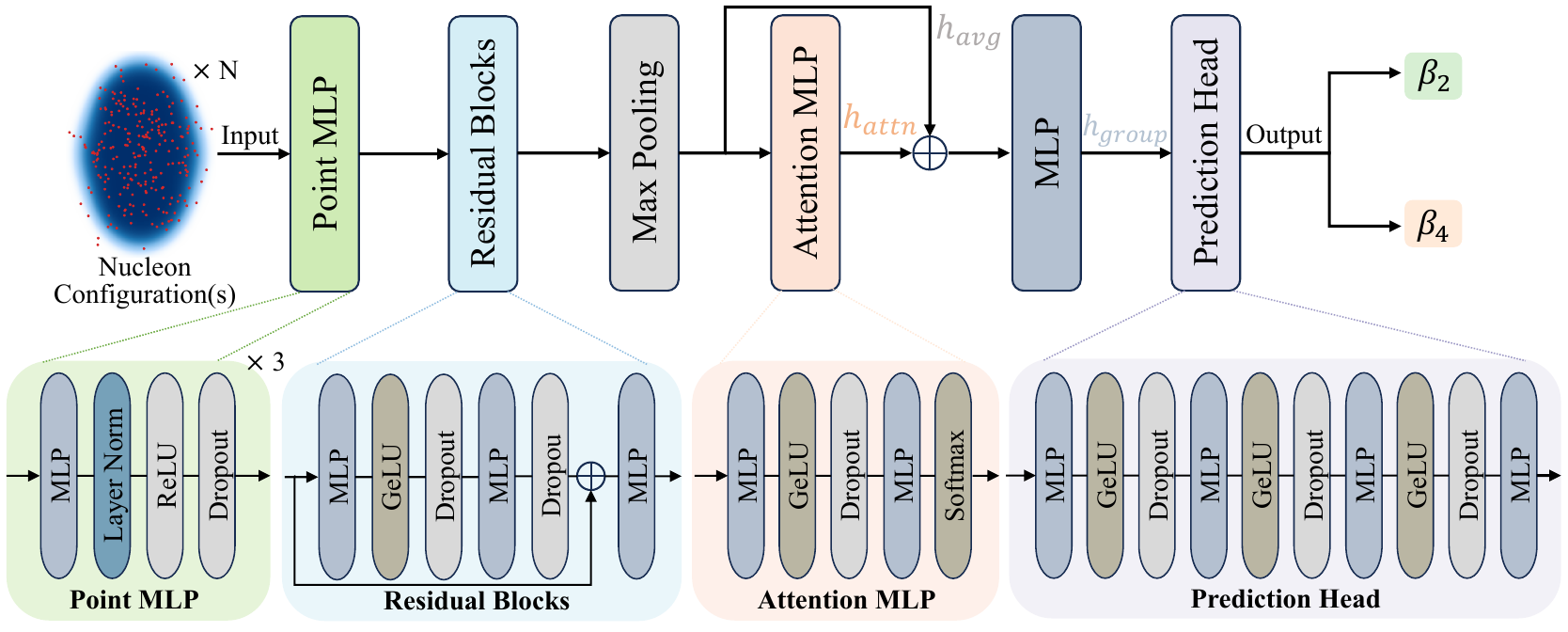}
		\caption{\label{network1}The structure of the point-cloud neural network used for inferring nuclear deformation parameters.}
	\end{figure*}
	
	As a baseline, we first investigate whether the deformation parameters $\beta_{2}$ and $\beta_{4}$ can be inferred directly from microscopic nucleon configurations. Nucleon positions are sampled from the deformed Woods--Saxon distribution,
	\begin{equation} \rho(r,\theta)=\frac{\rho_0}{1+\exp\!\left[\frac{r-R_0(1+\beta_2Y_{20}(\theta)+\beta_4Y_{40}(\theta))}{a}\right]}, \end{equation}
	where $\rho_{0}$ is the normalization constant, $R_{0}$ is the spherical Woods--Saxon radius, $\beta_{2}$ and $\beta_{4}$ are the quadrupole and hexadecapole deformation parameters, $Y_{20}$ and $Y_{40}$ are the associated spherical harmonics and $a$ is the surface diffuseness. For each $(\beta_{2},\beta_{4})$, nucleon coordinates are obtained via acceptance--rejection sampling in spherical coordinates: trial points $(r,\theta,\phi)$ are drawn uniformly up to a deformation-dependent cutoff $R_{\max}=R_{0}(1+0.63|\beta_{2}|+0.85|\beta_{4}|)+10a$ and accepted with probability $[1+\exp((r-R_{0}(1+\beta_{2}Y_{20}+\beta_{4}Y_{40}))/a)]^{-1}$, after which they are transformed to Cartesian coordinates; no additional hard-core repulsion is applied. For ${}^{238}\mathrm{U}$, we sample $\beta_{2}\in[-0.5,0.5]$ and $\beta_{4}\in[-0.2,0.2]$ on a $21\times21$ grid and generate 2000 independent configurations for each parameter set (grid point), each containing positions of all 238 nucleons. To mitigate fluctuations inherent to a single configuration, multiple events with identical deformation parameters are grouped into one training sample, with group size varied as $N=1, 5, 10$ and $20$ to probe the effect of statistical averaging on the deformation identifiability.
	
	The point-cloud neural network (see Fig.~\ref{network1}) \cite{Qi2017, Guo:2023zfk} is designed to be permutation-invariant with respect to input, following the general principles of set-based architectures \cite{Zaheer2017} and attention-based multiple instance learning \cite{Ilse2018}. In our study, as input, each nucleon configuration is first processed by a point-level multilayer perceptron that embeds the three-dimensional coordinates $(x,y,z)$ of each nucleon into a high-dimensional representation through three fully connected layers of widths 128, 256, and 512, each followed by LayerNorm \cite{Ba2016}, ReLU \cite{Maas2013} activation, and dropout with a rate of 0.1. Two residual blocks \cite{He2015} with a hidden width of 512 and GELU \cite{Dan2023} activations are then applied to further refine the local features. A max-pooling operation is performed across all nucleons within a configuration to obtain a single 512-dimensional feature vector representing the overall configuration.
	
	After configuration pooling, the network aggregates information from multiple configurations that share the same deformation parameters through two complementary branches. In the first branch, each configuration's feature vector with dimension 512 is processed by a small attention block composed of two fully connected layers (512 $\rightarrow$ 256 $\rightarrow$ 1) with GELU activation and dropout with rate of 0.1. The scalar outputs from this branch are normalized by a softmax activation over all $N$ configurations within the group to obtain attention weights $\alpha_i$ ($i=1,\ldots,N$), which quantify the relative importance of each configuration. These weights are then used to compute a weighted sum of the configuration features,
	\begin{equation}
		h_{\text{attn}} = \sum_{i=1}^{N} \alpha_i f_i,
	\end{equation}
	where $f_i \in \mathbb{R}^{512}$ represents the feature vector of the $i$th configuration. In parallel, the second branch directly averages the same configuration features without attention weighting,
	\begin{equation}
		h_{\text{avg}} = \frac{1}{N}\sum_{i=1}^{N} f_i,
	\end{equation}
	which preserves global information about the ensemble while suppressing event-by-event fluctuations. The two outputs $h_{\text{attn}}$ and $h_{\text{avg}}$ are concatenated and projected through a linear transformation (1024 $\rightarrow$ 512) to form the final group-level representation $h_{\text{group}}$. Finally, $h_{\text{group}}$ is passed to a regression head with four fully connected layers (512 $\rightarrow$ 256 $\rightarrow$ 128 $\rightarrow$ 64 $\rightarrow$ 2) using GELU activations and dropout with rate of 0.2 (the last layer is linear), which outputs the predicted deformation parameters $\beta_2$ and $\beta_4$. This design allows the model to combine the adaptivity of attention weighting with the stability of uniform averaging, yielding a compact 512-dimensional embedding that effectively captures both the common structure and the residual variation among configurations.
	
	All models are implemented in PyTorch \cite{Paszke2019} and trained in a supervised fashion using the Huber loss,
	\begin{equation}
		L_{\delta}(y,\hat{y}) =
		\begin{cases}
			\frac{1}{2}(y - \hat{y})^{2}, & \text{if } |y - \hat{y}| \le \delta, \\
			\delta \left( |y - \hat{y}| - \frac{1}{2}\delta \right), & \text{otherwise,}
		\end{cases}
	\end{equation}
	where $y$ is the true value, $\hat{y}$ is the network prediction and $\delta$ is a positive threshold controlling the transition between the quadratic and linear regimes. We set $\delta=1.0$ throughout this work. Optimization is performed using the Adam \cite{Kingma2017} optimizer with a learning rate of $3\times10^{-4}$ and a batch-wise training strategy. Model selection is based on validation loss, with 80\% of the $21\times21\times2000$ samples used for training and the remaining 20\% for validation; the selected models are then evaluated on an independent, higher-resolution test set consisting of $41\times41\times200$ samples. Performance is reported using both the Huber loss and the coefficient of determination ($R^2$) for $\beta_2$ and $\beta_4$, providing a quantitative baseline for how well the nuclear deformation information can be inferred directly from nucleon configurations before moving to more realistic initial-state modeling.
	
	\begin{figure*}[htbp]
		\centering
		\includegraphics[scale=0.5]{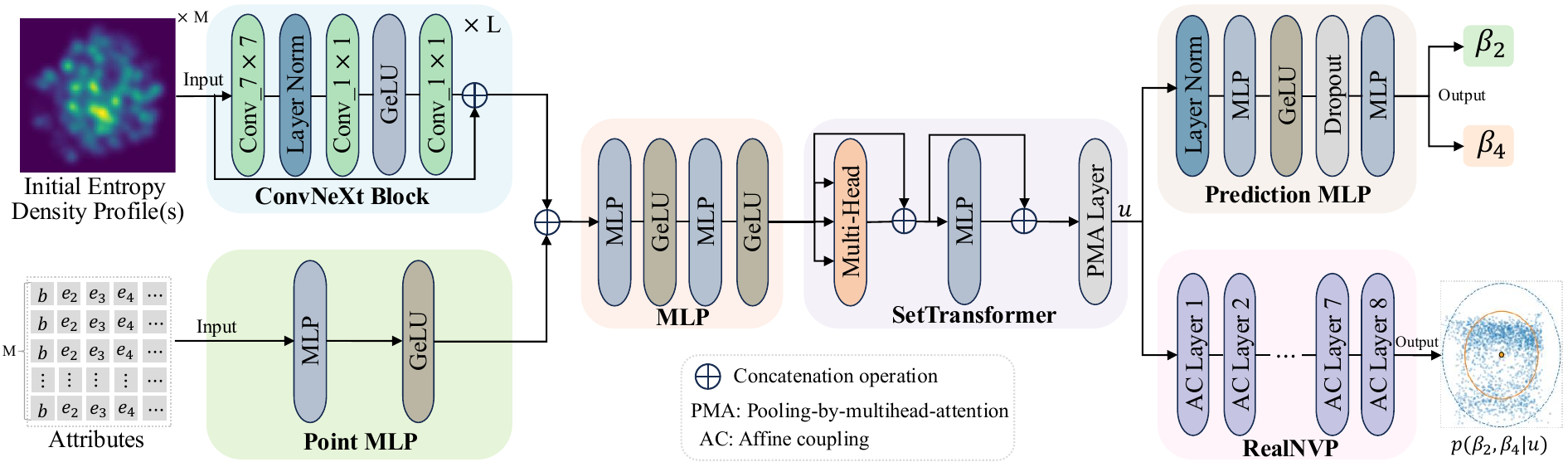}
		\caption{\label{network2}The structure of the neural network for regression and SBI from initial entropy density profiles. We use the MLP head for regression and the RealNVP head for SBI.}
	\end{figure*}	
	\subsection{Neural network for entropy density profiles}	
	We then investigate whether nuclear deformation parameters can be recovered from the initial entropy density profiles, with detailed network architecture depicted in Fig.~\ref{network2}. 
    
    For each pair of $\beta_2 \in [-0.5, 0.5]$ and $\beta_4 \in [-0.2, 0.2]$ on the $21 \times 21$ grid, T\raisebox{-0.5ex}{R}ENTo is used to simulate $^{238}$U+$^{238}$U collisions at $\sqrt{s_{NN}}=193$~GeV to generate the corresponding initial entropy density profile, producing 2000 independent events for each parameter pair in the selected centrality bins: 0--10\%, 30--40\%, 60--70\%, 90--100\% and 0--100\% (Centrality is defined according to multiplicity). Each event provides both a two-dimensional entropy density distribution of size $100 \times 100$ and a set of seven global attributes (impact parameter, eccentricities $e_{2}$--$e_{5}$, charged-particle multiplicity and participant number). These attributes complement the spatial information of the entropy maps by providing global geometric and participant properties. To suppress statistical fluctuations, multiple events with the same deformation parameters are grouped into one bag and each bag is treated as a single input sample. The bag size $M$ is varied as 1, 10, 50 and 100, so that the network can learn from ensembles with different levels of statistical averaging. This grouping is crucial, since the stochastic nature of nucleon dynamics and entropy deposition strongly blurs the deformation signal within a single event. Even after averaging over 100 events, the entropy-density profiles for different $(\beta_2,\beta_4)$ values remain visually very similar; representative examples are provided in Appendix~\ref{appendixA}.
	
	For each bag, the entropy density maps are processed by a convolutional backbone network. In this study we employ ConvNeXt-Small \cite{Liu2022} for both the regression and the SBI setups. ConvNeXt is a modern convolutional architecture inspired by the design philosophy of vision transformers, featuring large kernel sizes, inverted bottlenecks, GELU activations and extensive use of LayerNorm in place of BatchNorm. It has been shown to achieve state-of-the-art performance on large-scale image benchmarks while retaining the computational efficiency, inductive biases and stability of classic CNNs. These properties make ConvNeXt particularly suitable for modeling fluctuating initial-state profiles in heavy-ion collisions. Each event image is encoded into a 128-dimensional feature vector, denoted as $f_i \in \mathbb{R}^{128}$.
	
	In parallel, the seven per-event attributes are processed by a small fully connected network (multilayer perceptron, MLP) consisting of a single linear layer followed by a GELU activation, producing a 128-dimensional attribute-embedding $a_i \in \mathbb{R}^{128}$. The image feature $f_i$ and the attribute-embedding $a_i$ are concatenated and passed through a two-layer MLP (linear–GELU–linear–GELU) to yield a 128-dimensional event-level embedding $z_i \in \mathbb{R}^{128}$. For a bag containing $M$ events, the encoder outputs a set $\{z_i\}_{i=1}^{M}$ with shape $[M,128]$, which is then aggregated into a single bag-level representation through a Set Transformer \cite{Lee2019}. In preliminary tests, we also considered directly stacking the $M$ event images along the channel dimension and feeding them into a wider CNN for joint processing. However, this strategy performed substantially worse than the present design based on single-event encoding followed by Set Transformer aggregation, likely because it does not explicitly preserve the set structure or permutation invariance of the events within each bag.
	
	The Set Transformer aggregates variable-sized inputs through self-attention and permutation-invariant pooling. Our implementation consists of two stacked self-attention blocks (SAB) with hidden dimension 256 and four attention heads, followed by a pooling-by-multihead-attention (PMA) layer with a single seed vector. The PMA output is linearly projected to a 256-dimensional bag-level representation $u \in \mathbb{R}^{256}$ that summarizes the global structure of all events within the bag. On top of this representation, two inference strategies, hereafter referred to as the regression and SBI setups, are considered. In the regression setup, a small multilayer perceptron directly outputs point estimates of $\beta_{2}$ and $\beta_{4}$, trained with the Huber loss and evaluated with the coefficient of determination ($R^2$). The regression head consists of LayerNorm applied to $u$, followed by a linear layer expanding to 512 units ($2U_{\text{dim}}$), a GELU activation, dropout (0.1) and a final linear layer projecting to two outputs corresponding to $\beta_{2}$ and $\beta_{4}$. This lightweight structure provides an efficient mapping from the bag-level representation to the deformation parameters. In the SBI setup, the regression head is replaced by a conditional normalizing flow, specifically RealNVP \cite{Dinh2017}, which models the full posterior distribution $p(\beta_{2},\beta_{4}\,|\,u)$. The RealNVP flow is implemented with eight affine coupling layers, each using a hidden dimension of 128, providing a flexible yet tractable conditional density estimator conditioned on the 256-dimensional context $u$.
	
	All models are implemented in PyTorch and trained with the AdamW \cite{Ilya2019} optimizer using a cosine learning-rate schedule with warmup \cite{Ilya2017}. Model selection is based on validation performance, with 80\% of the $21\times21\times2000$ samples used for training and the remaining 20\% for validation; the selected models are then evaluated on an independent, higher-resolution test set consisting of $41\times41\times200$ samples. For both the regression and SBI setups, we adopt ConvNeXt-Small as the backbone, with a learning rate of $1\times10^{-5}$ and a weight decay of $1\times10^{-6}$.

	\section{Results}\label{Sec3}
	
	\begin{figure*}[htbp]
		\centering
		\includegraphics[scale=0.8]{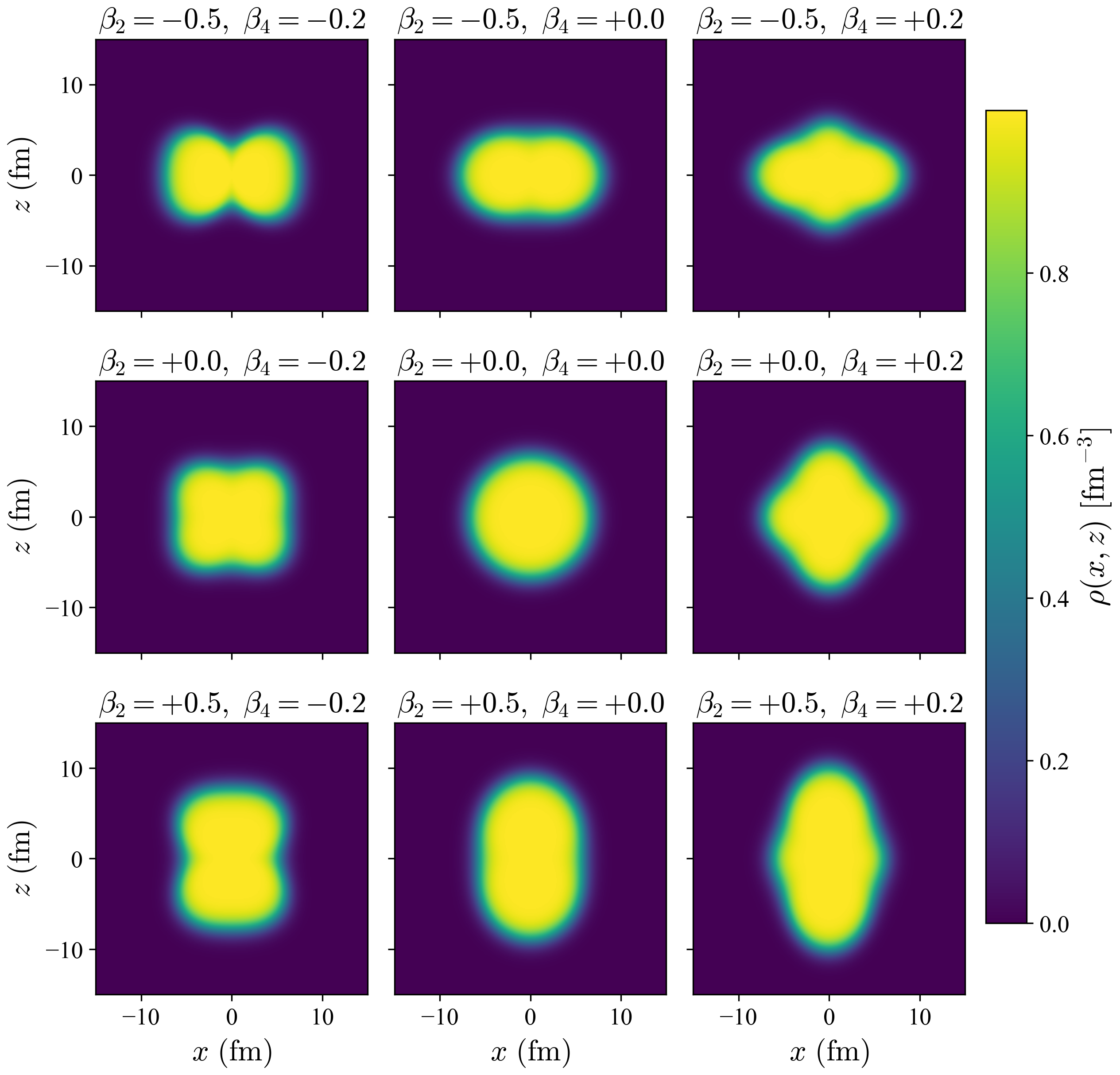}
		\caption{\label{DWS}Typical deformed Woods--Saxon distributions for representative deformation parameters of $(\beta_{2},\beta_{4})$ in the $x$--$z$ plane.}
	\end{figure*}
	
	We begin by illustrating the impact of the deformation parameters $\beta_{2}$ and $\beta_{4}$ on the underlying nucleon density profiles. Fig.~\ref{DWS} shows representative deformed Woods--Saxon distributions in the $x$--$z$ plane. The quadrupole deformation $\beta_{2}$ controls the global elongation of the nucleus, producing oblate shapes for $\beta_{2}<0$ and prolate shapes for $\beta_{2}>0$. In contrast, the hexadecapole deformation $\beta_{4}$ induces more localized modifications to the surface curvature, generating box-like distortions when negative and diamond-like distortions when positive, while leaving the overall size largely unaffected. These patterns demonstrate that $\beta_{2}$ and $\beta_{4}$ imprint distinct geometric signatures on the nuclear profile and can, in principle, be learned by machine-learning models. Yet it is worth noting that Fig.~\ref{DWS} shows the smooth deformed nucleon distribution, whereas actual heavy-ion collisions involve fluctuating nucleon configurations sampled from this underlying distribution.
	
	Physically, the difference in recoverability between the two parameters arises from their relative impact on the global geometry as depicted in Fig.~\ref{DWS}. The quadrupole deformation $\beta_{2}$ strongly influences the leading ellipticity $\varepsilon_{2}$ of the initial-state, producing a large and robust geometric signal, while $\beta_{4}$ affects higher-order moments such as $\varepsilon_{4}$ and subleading corrections to $\varepsilon_{2}$, whose amplitudes are smaller and more easily overwhelmed by stochastic nucleon fluctuations, making $\beta_{4}$ intrinsically harder to extract than $\beta_{2}$.
	
	\subsection{Identifiability of $(\beta_{2}, \beta_{4})$ from nucleon configurations}
	
	\begin{figure}[htbp]
		\centering
		\includegraphics[scale=0.6]{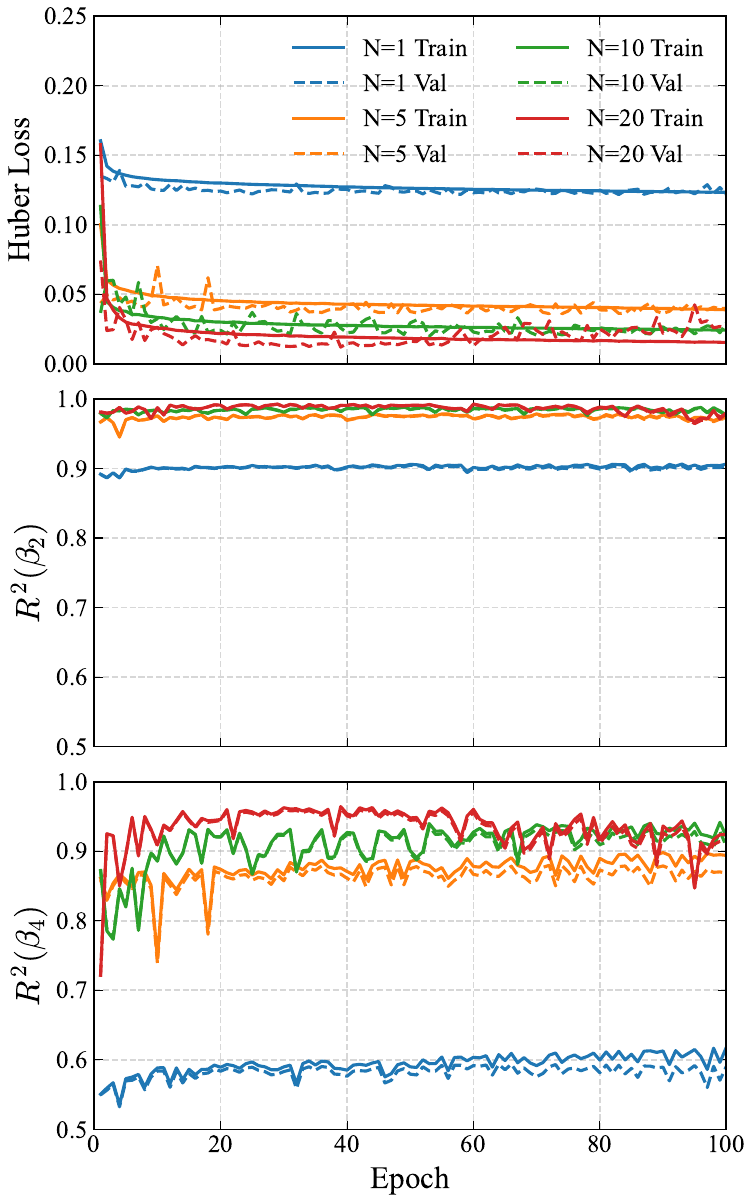}
		\caption{\label{Config_Train_Results}Training and validation performance of the point-cloud model for nucleon configurations at different group sizes $N$.}
	\end{figure}
	
	\begin{table*}[!htb]
		\renewcommand{\arraystretch}{1.5}
		\centering
		\caption{Best $R^2$ scores of the point-cloud model for different group sizes $N$ when regressing $\beta_2$ and $\beta_4$ from nucleon configurations.}
		\begin{tabular}{>{\centering\arraybackslash}p{1.2cm}
				>{\centering\arraybackslash}p{2.5cm}
				>{\centering\arraybackslash}p{2.5cm}
				>{\centering\arraybackslash}p{2.5cm}
				>{\centering\arraybackslash}p{2.5cm}
				>{\centering\arraybackslash}p{2.5cm}
				>{\centering\arraybackslash}p{2.5cm}}
			\hline\hline
			\textbf{$N$} & 
			\textbf{Train $R^2$ ($\beta_2$)} & 
			\textbf{Train $R^2$ ($\beta_4$)} & 
			\textbf{Val $R^2$ ($\beta_2$)} & 
			\textbf{Val $R^2$ ($\beta_4$)} & 
			\textbf{Test $R^2$ ($\beta_2$)} &
			\textbf{Test $R^2$ ($\beta_4$)} \\
			\hline
			1  & 0.90543 & 0.61228 & 0.90296 & 0.59229 & 0.89969 & 0.58252 \\
			5  & 0.97778 & 0.89338 & 0.97671 & 0.87934 & 0.97577 & 0.87686 \\
			10 & 0.98639 & 0.93987 & 0.98572 & 0.93561 & 0.98528 & 0.93194 \\
			20 & 0.99076 & 0.96428 & 0.99062 & 0.96141 & 0.99062 & 0.95998 \\
			\hline\hline
		\end{tabular}
		\label{best_config}
	\end{table*}
	
	\begin{figure*}[htbp]
		\centering
		\includegraphics[scale=0.45]{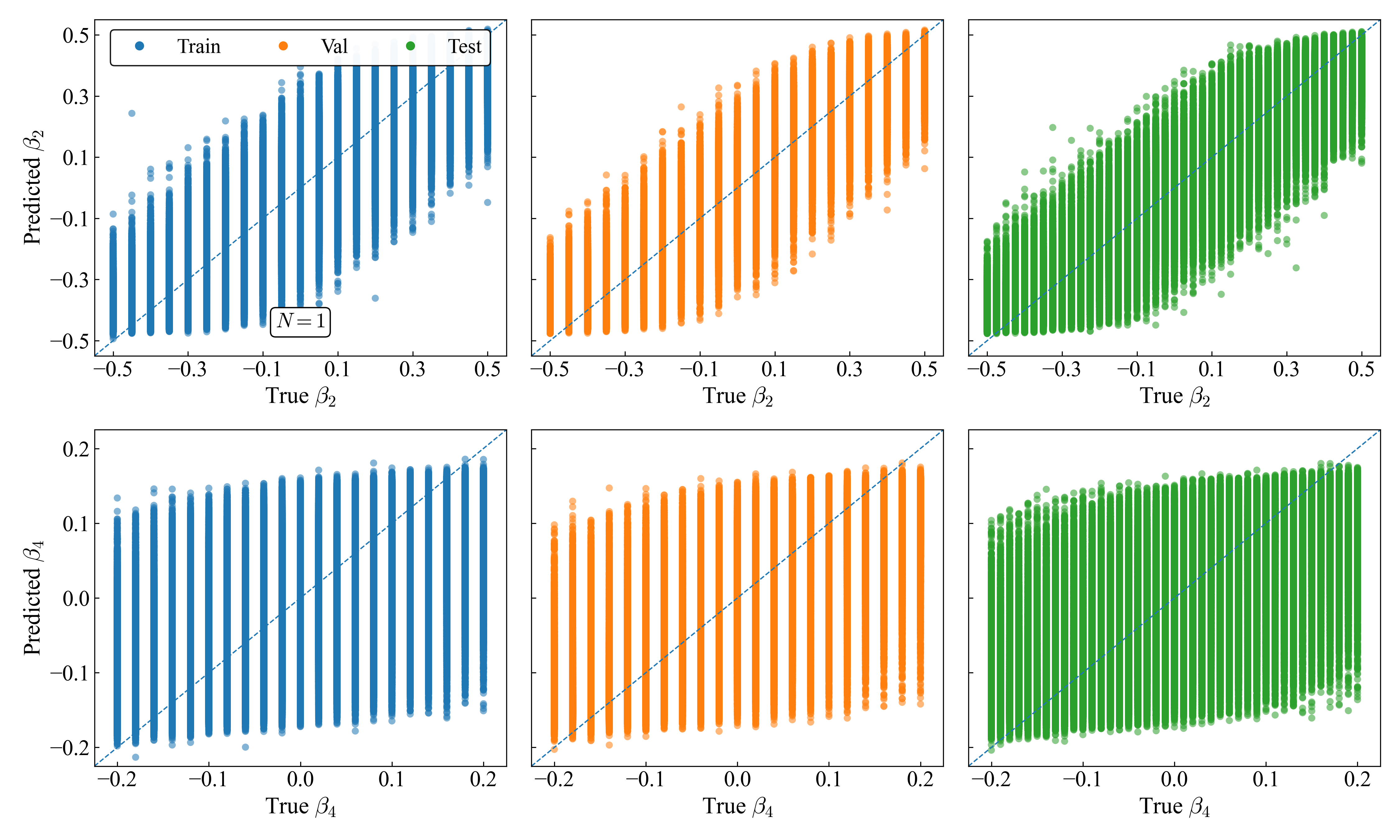}
		\caption{\label{config_scatter_N1}Predicted versus true $\beta_{2}$ (top) and $\beta_{4}$ (bottom) for nucleon-configuration regression at $N=1$. The diagonal denotes perfect agreement.}
	\end{figure*}
	
	\begin{figure*}[htbp]
		\centering
		\includegraphics[scale=0.45]{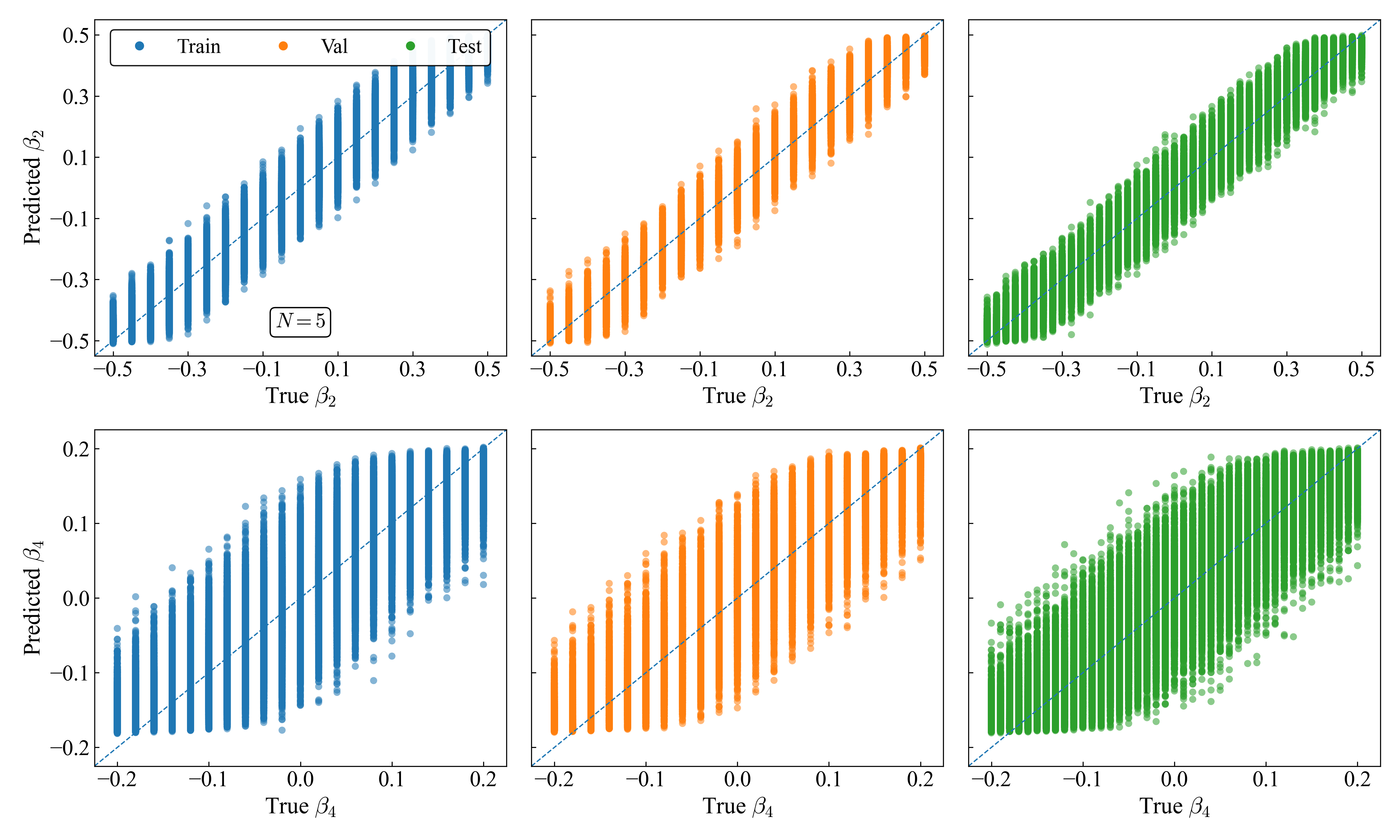}
		\caption{\label{config_scatter_N5}Same as Fig.~\ref{config_scatter_N1}, but for $N=5$.}
	\end{figure*}
	
	\begin{figure*}[htbp]
		\centering
		\includegraphics[scale=0.45]{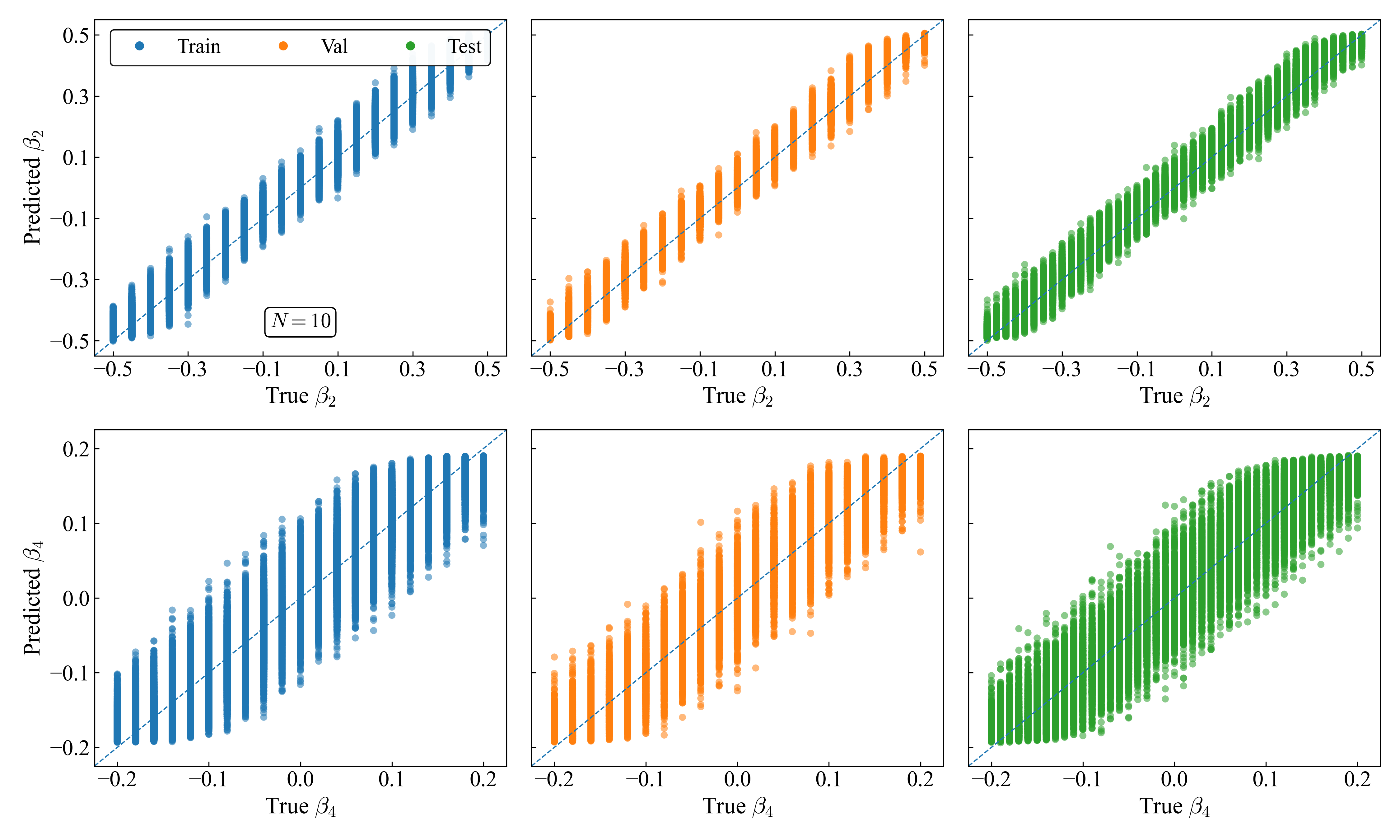}
		\caption{\label{config_scatter_N10}Same as Fig.~\ref{config_scatter_N1}, but for $N=10$.}
	\end{figure*}
	
	\begin{figure*}[htbp]
		\centering
		\includegraphics[scale=0.45]{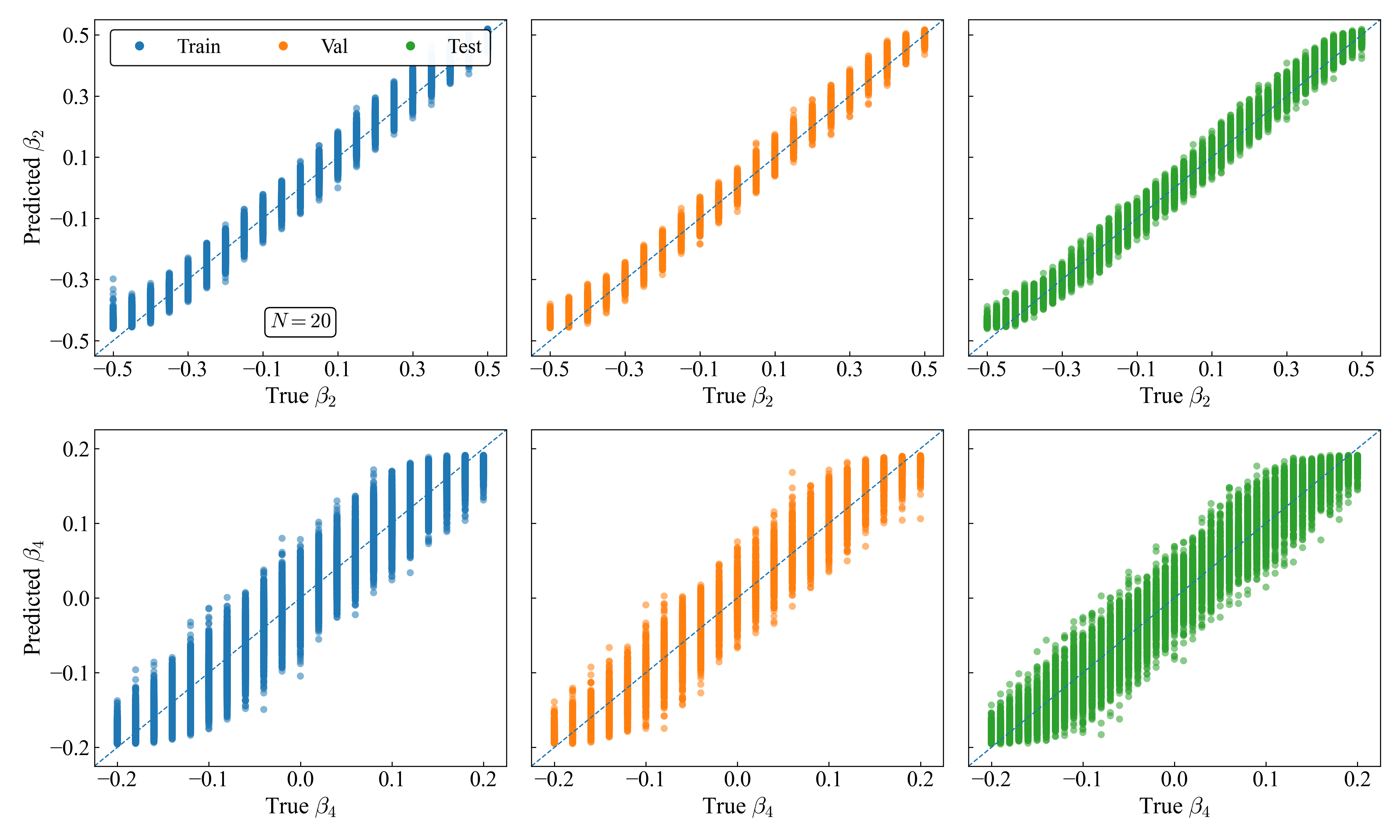}
		\caption{\label{config_scatter_N20}Same as Fig.~\ref{config_scatter_N1}, but for $N=20$.}
	\end{figure*}
	
	Figure~\ref{Config_Train_Results} summarizes the training performance of the point-cloud neural network for different event group sizes $N=1, 5, 10,$ and $20$. The top panel shows the Huber losses for the training and validation sets, which decrease rapidly during the first few epochs and then reach stable plateaus. The loss values systematically drop as $N$ increases, indicating that aggregating multiple configurations with identical deformation parameters significantly improves the model's stability and generalization. The middle and bottom panels display the coefficients of determination, $R^{2}(\beta_{2})$ and $R^{2}(\beta_{4})$, respectively. For single-event inputs ($N=1$), the model achieves only moderate correlations with the true deformation parameters, while for $N \ge 5$ the $R^{2}$ values quickly rise above 0.9 and remain stable throughout training. As $N$ increases from 5 to 20, the performance continues to improve but gradually saturates, suggesting that the information gained from additional events becomes redundant once the stochastic fluctuations are sufficiently suppressed. The inference of $\beta_{2}$ is generally more accurate and less fluctuating than that of $\beta_{4}$, reflecting the stronger imprint of quadrupole deformation on the nucleon geometry. Overall, these results demonstrate that grouping multiple events effectively suppresses statistical fluctuations and enables reliable extraction of deformation parameters from nucleon configurations.
	
	Table~\ref{best_config} summarizes the best $R^{2}$ scores of the point-cloud neural network for different event group sizes $N$. The results show a clear improvement as $N$ increases, with the accuracy for $\beta_{2}$ consistently higher than that for $\beta_{4}$. When $N$ reaches 20, the $R^{2}$ values exceed 0.99 for $\beta_{2}$ and about 0.96 for $\beta_{4}$ across all data splits, indicating that the network achieves high predictive precision for both deformation parameters. Figs.~\ref{config_scatter_N1}, \ref{config_scatter_N5}, \ref{config_scatter_N10} and \ref{config_scatter_N20} further visualize the correlation between the predicted and true values. For $N=1$, the predicted parameters exhibit a wide dispersion and systematic deviation from the diagonal, particularly for $\beta_{4}$, reflecting the limited information can be extracted by single-event nucleon configurations because of fluctuations. As $N$ increases to 5 and 10, the scatter clouds become more compact and symmetric around the diagonal, with fewer outliers and reduced bias. The improvement is observed consistently across the training, validation and test sets, suggesting that the network generalizes well once multi-event information is introduced. At $N=20$, the predicted points for both $\beta_{2}$ and $\beta_{4}$ are tightly aligned along the one-to-one line, forming narrow bands that indicate strong linear correlation and minimal systematic deviation. A slightly larger residual spread remains for $\beta_{4}$, consistent with its weaker imprint on the underlying nuclear geometry. These results collectively confirm that combining multiple configurations effectively suppresses event-by-event fluctuations, enhances the robustness of deformation inference and that the performance gradually saturates once sufficient statistical averaging is achieved.

	\subsection{Results from entropy density profiles}
	
	In this work, we analyze five centrality bins: 0--10\%, 30--40\%, 60--70\%, 90--100\%, and 0--100\%. To keep the presentation concise, only the detailed results for the 0--10\% bin are shown in the main text, whereas representative results for the remaining centrality bins are included in Appendix~\ref{appendixB}. Additional image results are available from the authors upon request.
	
	\subsubsection{Regression}
	
	\begin{figure}[htbp]
		\centering
		\includegraphics[scale=0.6]{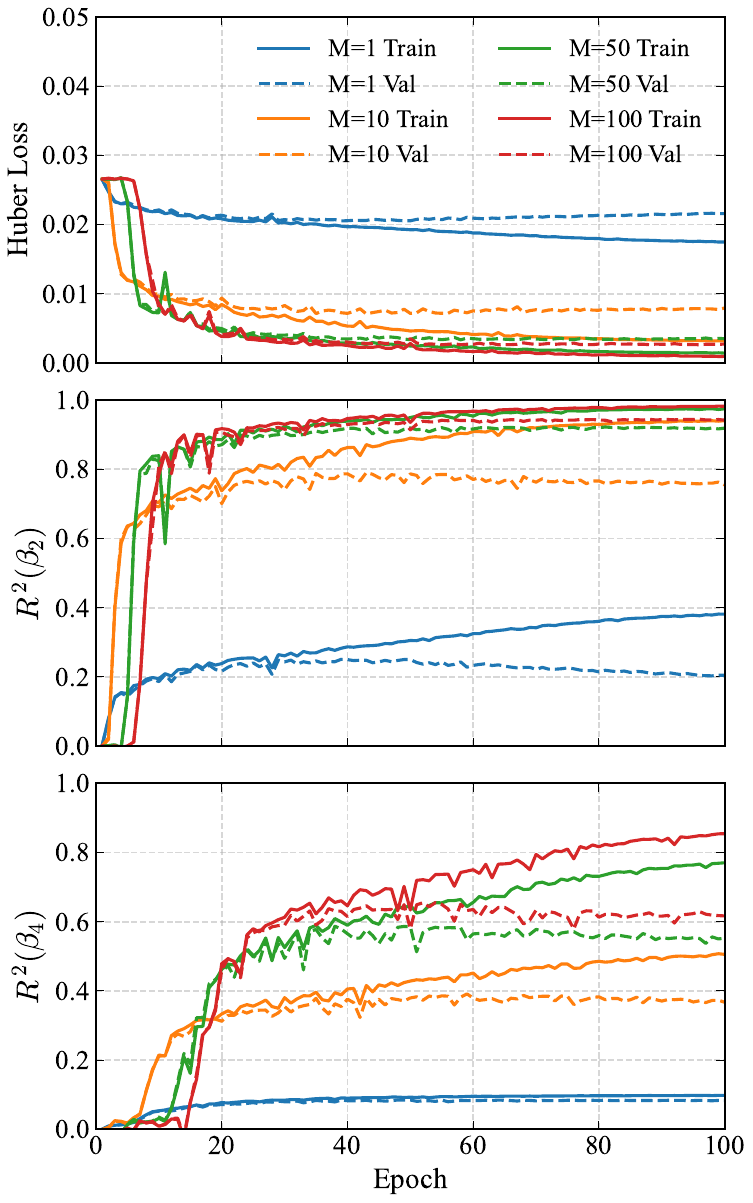}
		\caption{\label{REG_Train_Results}Training and validation curves of the regression model using T\raisebox{-0.5ex}{R}ENTo entropy profiles for different bag sizes $M$ (0--10\% centrality).}
	\end{figure}
	
	\begin{table*}[!htb]
		\renewcommand{\arraystretch}{1.5}
		\centering
		\caption{Best regression performance on T\raisebox{-0.5ex}{R}ENTo entropy profiles at different bag sizes $M$, showing the corresponding $R^2$ values for $\beta_2$ and $\beta_4$ (0–10\% centrality).}
		\begin{tabular}{>{\centering\arraybackslash}p{1.2cm}
				>{\centering\arraybackslash}p{2.5cm}
				>{\centering\arraybackslash}p{2.5cm}
				>{\centering\arraybackslash}p{2.5cm}
				>{\centering\arraybackslash}p{2.5cm}
				>{\centering\arraybackslash}p{2.5cm}
				>{\centering\arraybackslash}p{2.5cm}}
			\hline\hline
			\textbf{$M$} & 
			\textbf{Train $R^2$ ($\beta_2$)} & 
			\textbf{Train $R^2$ ($\beta_4$)} & 
			\textbf{Val $R^2$ ($\beta_2$)} & 
			\textbf{Val $R^2$ ($\beta_4$)} & 
			\textbf{Test $R^2$ ($\beta_2$)} &
			\textbf{Test $R^2$ ($\beta_4$)} \\
			\hline
			1   & 0.28327 & 0.08882 & 0.25559 & 0.08598 & 0.24239 & 0.08706 \\
			10  & 0.87299 & 0.41458 & 0.79058 & 0.37849 & 0.77968 & 0.36816 \\
			50  & 0.95940 & 0.67016 & 0.92116 & 0.58417 & 0.91914 & 0.56038 \\
			100 & 0.96445 & 0.74083 & 0.94173 & 0.64976 & 0.93200 & 0.61965 \\
			\hline\hline
		\end{tabular}
		\label{best_reg}
	\end{table*}
	
	\begin{figure*}[htbp]
		\centering
		\includegraphics[scale=0.45]{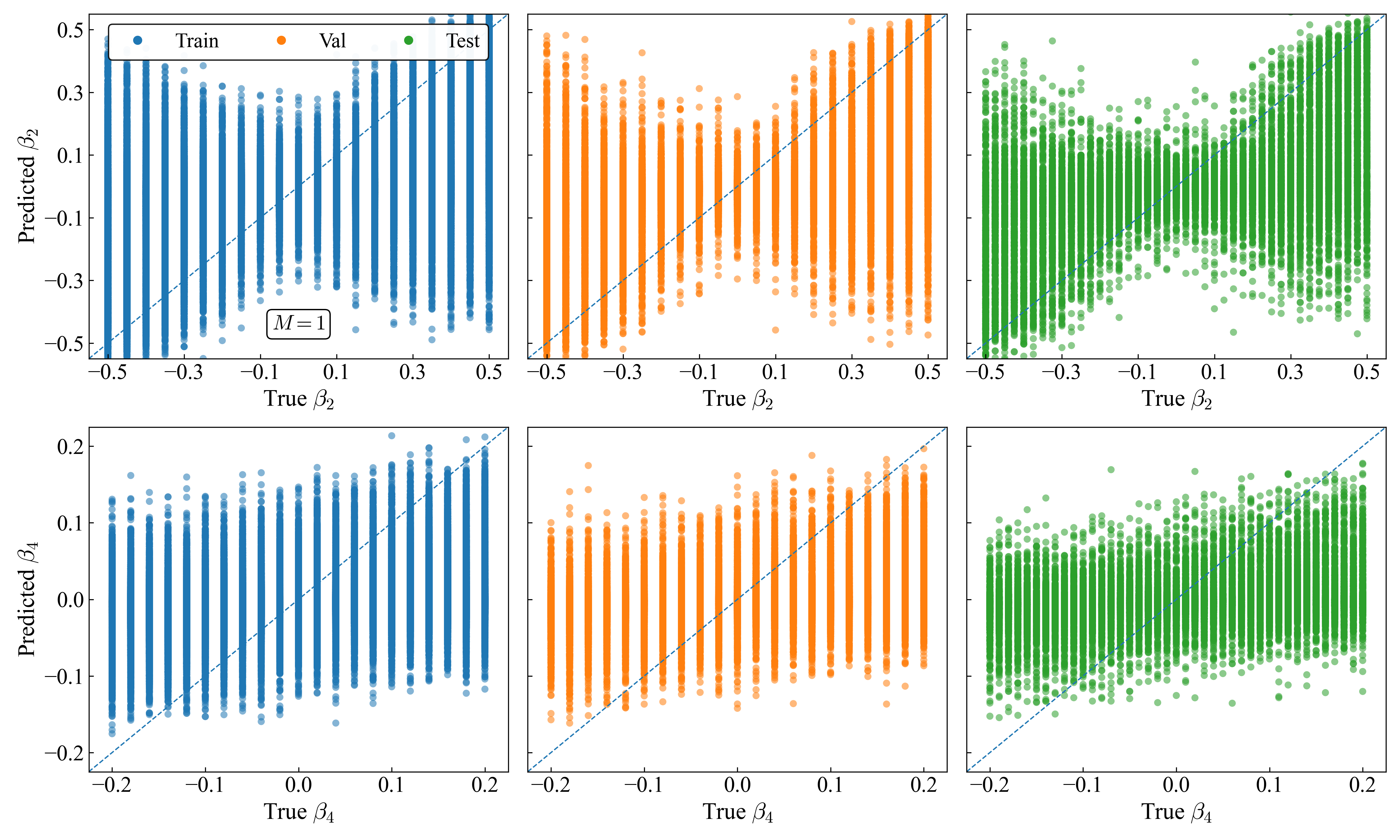}
		\caption{\label{reg_scatter_M1}Predicted versus true $\beta_{2}$ (top) and $\beta_{4}$ (bottom) for T\raisebox{-0.5ex}{R}ENTo regression at $M=1$ (0--10\% centrality). The diagonal denotes perfect agreement.}
	\end{figure*}
	
	\begin{figure*}[htbp]
		\centering
		\includegraphics[scale=0.45]{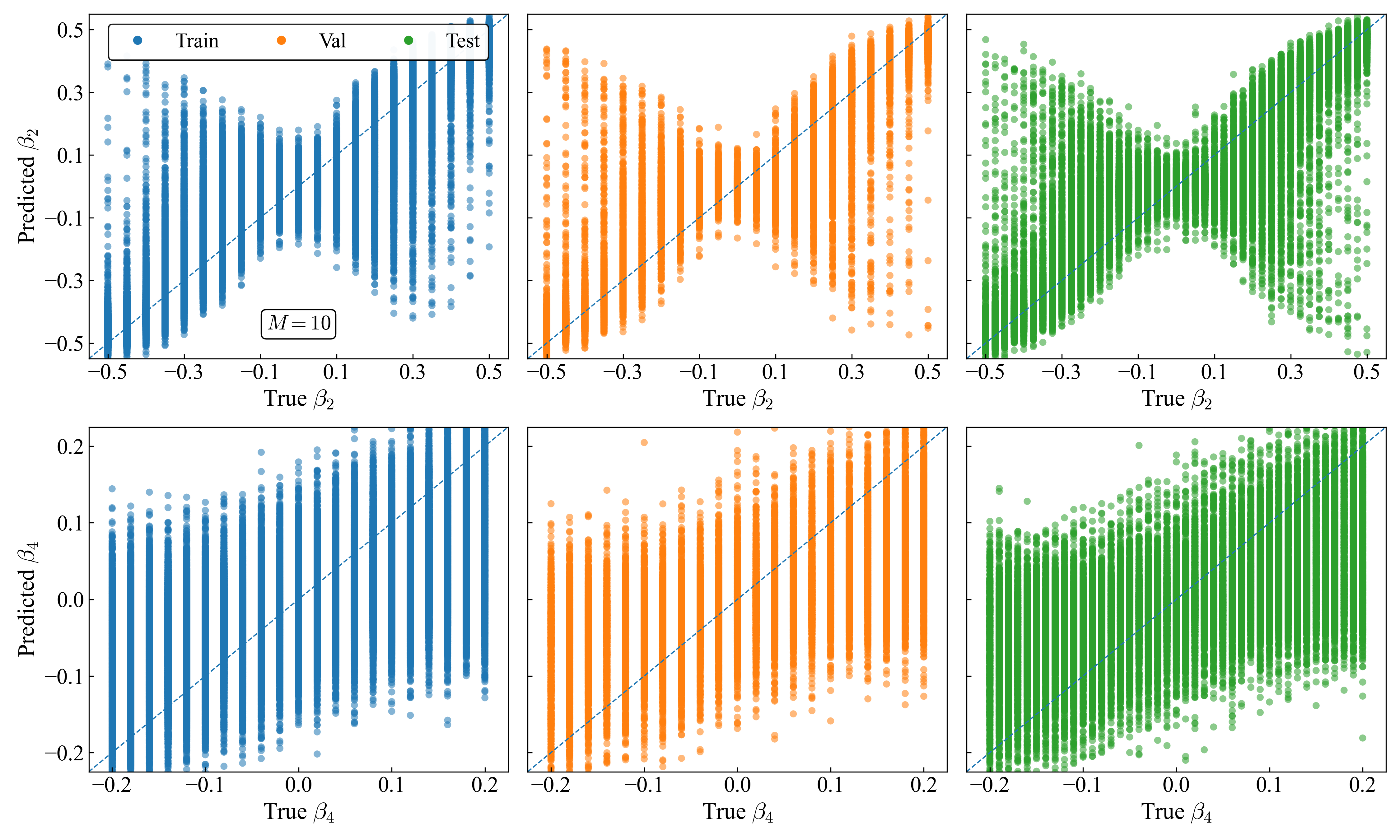}
		\caption{\label{reg_scatter_M10}Same as Fig.~\ref{reg_scatter_M1}, but for $M=10$ (0--10\% centrality).}
	\end{figure*}
	
	\begin{figure*}[htbp]
		\centering
		\includegraphics[scale=0.45]{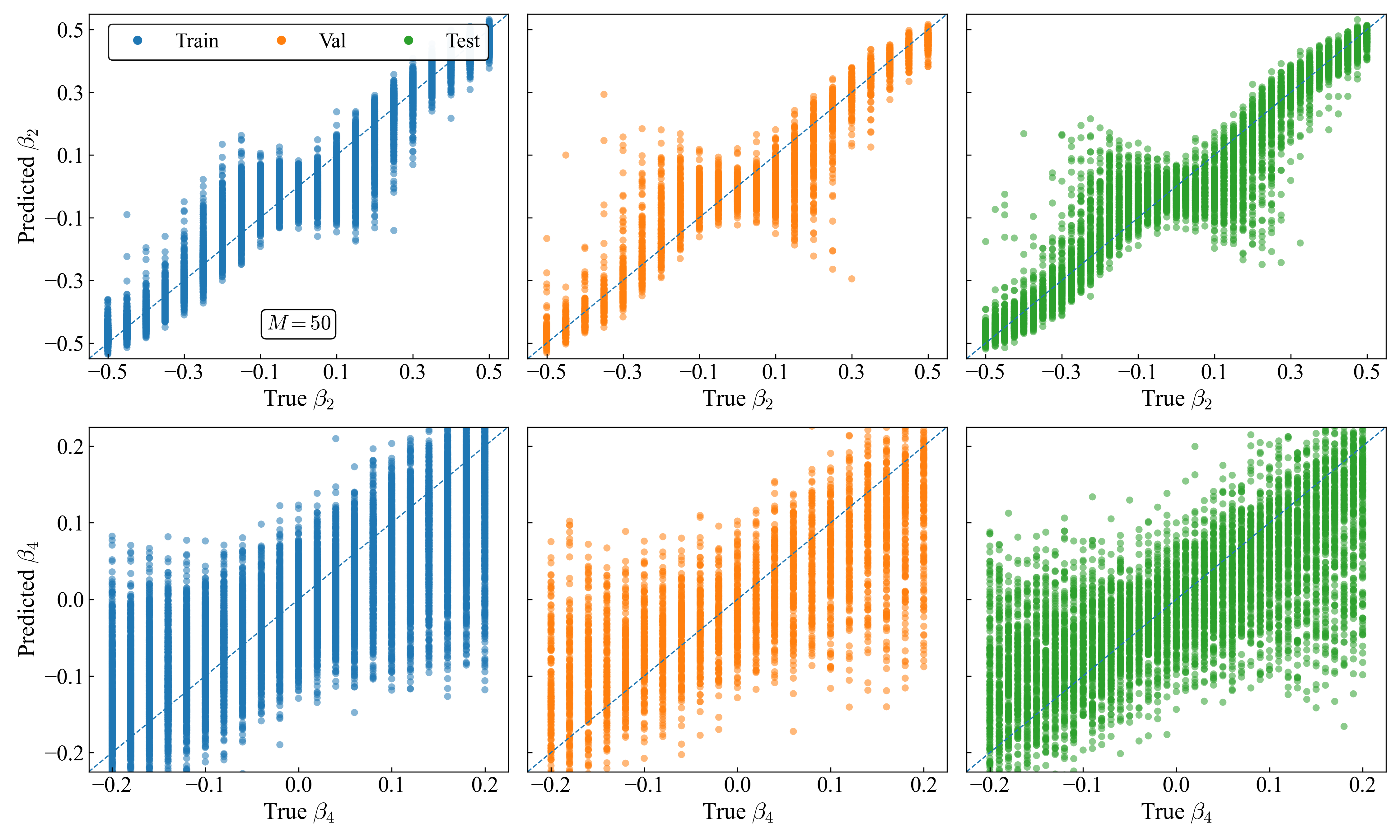}
		\caption{\label{reg_scatter_M50}Same as Fig.~\ref{reg_scatter_M1}, but for $M=50$ (0--10\% centrality).}
	\end{figure*}
	
	\begin{figure*}[htbp]
		\centering
		\includegraphics[scale=0.45]{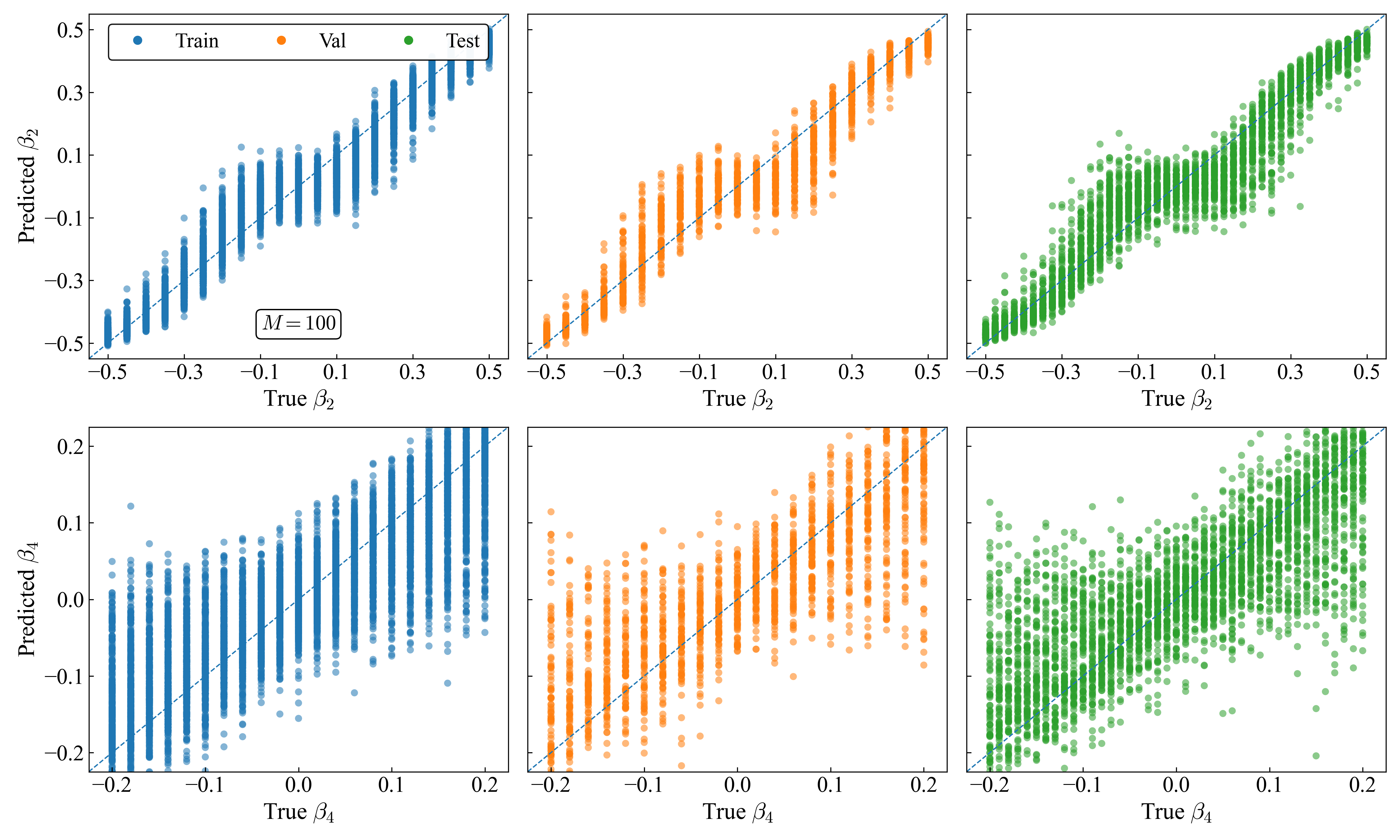}
		\caption{\label{reg_scatter_M100}Same as Fig.~\ref{reg_scatter_M1}, but for $M=100$ (0--10\% centrality).}
	\end{figure*}
	
	Figure~\ref{REG_Train_Results} shows the training and validation performance of the regression model for T\raisebox{-0.5ex}{R}ENTo entropy-density profiles with bag sizes $M=1,10,50,$ and $100$ at 0--10\% centrality. The validation Huber loss (top panel) decreases noticeably faster and reaches lower plateaus as $M$ increases, reflecting the suppression of event-level fluctuations through multi-event averaging. In contrast, the training loss continues to decrease even after the validation loss saturates, indicating mild overfitting that becomes more pronounced for larger $M$. The validation $R^{2}$ scores (middle and bottom panels) exhibit the same trend: with $M=1$, the deformation signal is too weak to be captured reliably, whereas larger bags yield smoother and substantially improved validation performance. Increasing $M$ therefore enhances the stability of training and the overall learnability of the deformation parameters.
	
	The best regression performance for the 0--10\% centrality is summarized in Tab.~\ref{best_reg} and further illustrated in Figs.~\ref{reg_scatter_M1}, \ref{reg_scatter_M10}, \ref{reg_scatter_M50} and \ref{reg_scatter_M100}. As the bag size increases, deformation reconstruction improves noticeably: at $M=1$, the predictions for both $\beta_{2}$ and $\beta_{4}$ show large scatter and strong deviations from the identity line, reflecting the weakness of deformation information in single-event entropy profiles. At $M=10$, a clearer monotonic trend appears but still remains noisy, especially for $\beta_{4}$. For $M=50$ and $M=100$, the points cluster tightly around the diagonal, with $\beta_{2}$ achieving near-linear recovery and $\beta_{4}$ showing a compressed but still well-defined correlation. These behaviors are consistent with the geometric roles of the two parameters: $\beta_{2}$ governs global ellipticity and leaves a strong imprint, while $\beta_{4}$ affects only higher-order surface curvature and is more easily masked by fluctuations. Two characteristic features seen in the scatter plots echo previous findings \cite{Pang2019}: configurations with equal $|\beta_{2}|$ (prolate vs.\ oblate) are harder to distinguish for small $M$, producing X-shaped patterns, while the predictions tighten noticeably near $\beta_{2}=0$, where the nearly spherical geometry yields a simpler and more stable mapping. Overall, deformation information becomes progressively recoverable with increasing bag size, with $\beta_{2}$ consistently easier to infer than $\beta_{4}$.

	\subsubsection{Simulation-based inference}
	
	\begin{figure}[htbp]
		\centering
		\includegraphics[scale=0.6]{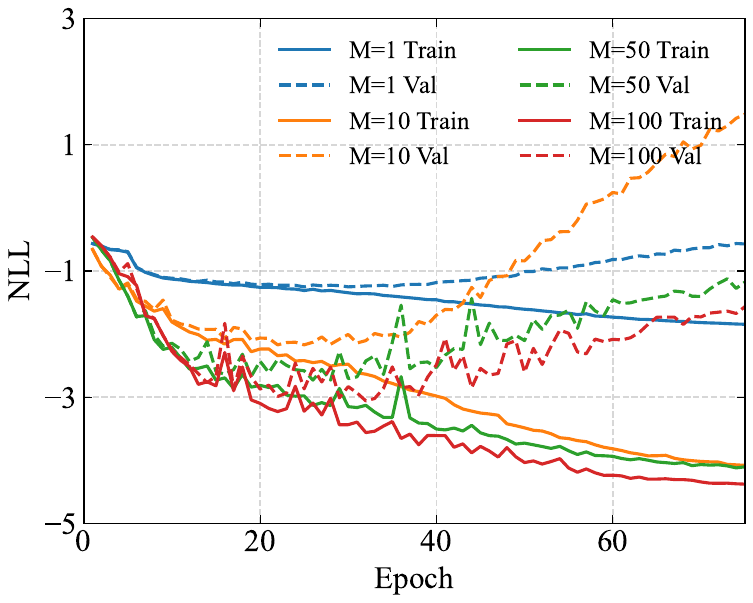}
		\caption{\label{SBI_Train_Results}Training and validation negative log-likelihood (NLL) of the SBI model at different bag sizes (0--10\% centrality).}
	\end{figure}
	
	\begin{table*}[!htb]
		\renewcommand{\arraystretch}{1.5}
		\centering
		\caption{Likelihood gains of the SBI model relative to a Gaussian baseline for different bag sizes $M$ (0–10\% centrality).}
		\begin{tabular}{>{\centering\arraybackslash}p{2.4cm}
				>{\centering\arraybackslash}p{5cm}
				>{\centering\arraybackslash}p{5cm}
				>{\centering\arraybackslash}p{5cm}}
			\hline\hline
			\textbf{$M$} & 
			\textbf{Train LL Gain (vs. Baseline)} & 
			\textbf{Val LL Gain (vs. Baseline)} & 
			\textbf{Test LL Gain (vs. Baseline)} \\
			\hline
			1   & 0.83934 & 0.79699 & 0.75596  \\
			10  & 1.87198 & 1.70161 & 1.67243  \\
			50  & 2.66984 & 2.25886 & 2.21312  \\
			100 & 3.09021 & 2.58878 & 2.40813  \\
			\hline\hline
		\end{tabular}
		\label{best_sbi}
	\end{table*}
	
	\begin{figure}[htbp]
		\centering
		\includegraphics[scale=0.6]{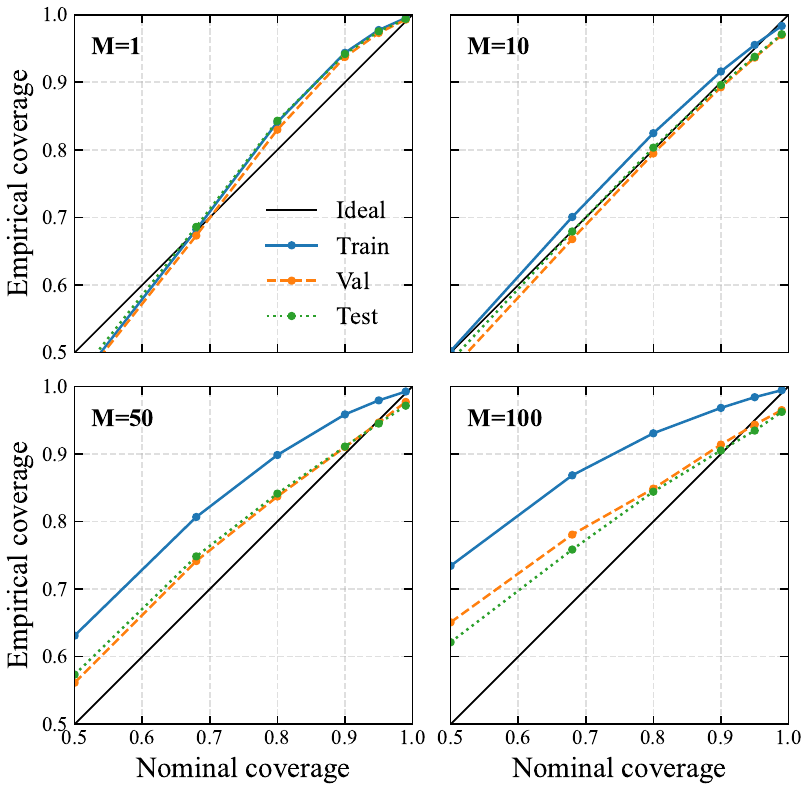}
		\caption{\label{SBI_Cov}Empirical coverage versus nominal coverage for the posterior distributions obtained from SBI at different bag sizes $M$ (0--10\% centrality).}
	\end{figure}
	
	\begin{figure}[htbp]
		\centering
		\includegraphics[scale=0.6]{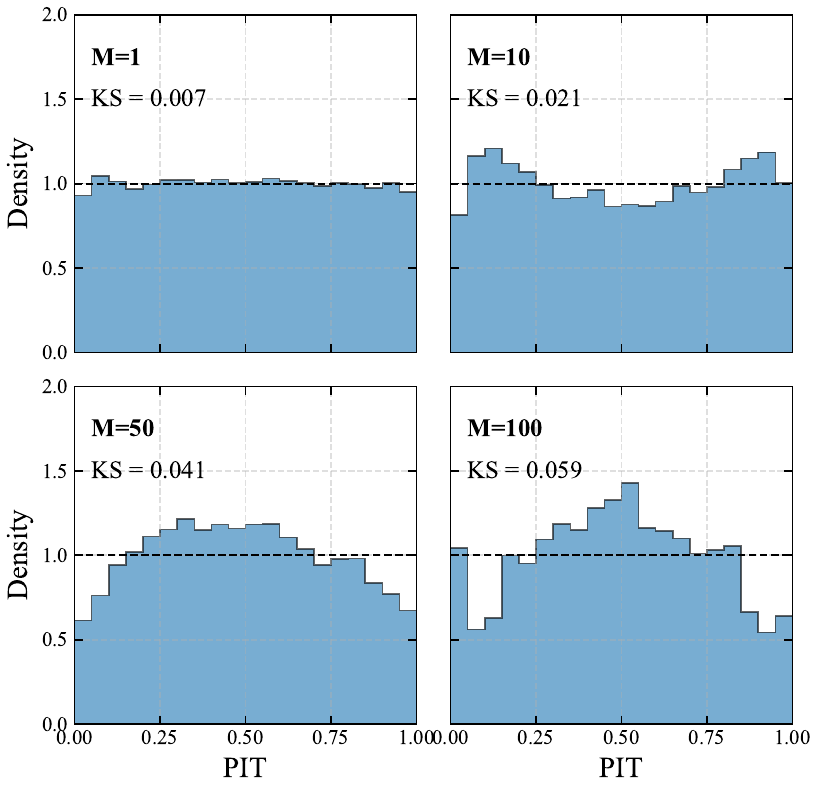}
		\caption{\label{SBI_PIT}Probability integral transform histograms for the posterior samples at different bag sizes $M$ (0--10\% centrality). }
	\end{figure}
	
	\begin{figure*}[htbp]
		\centering
		\includegraphics[scale=0.7]{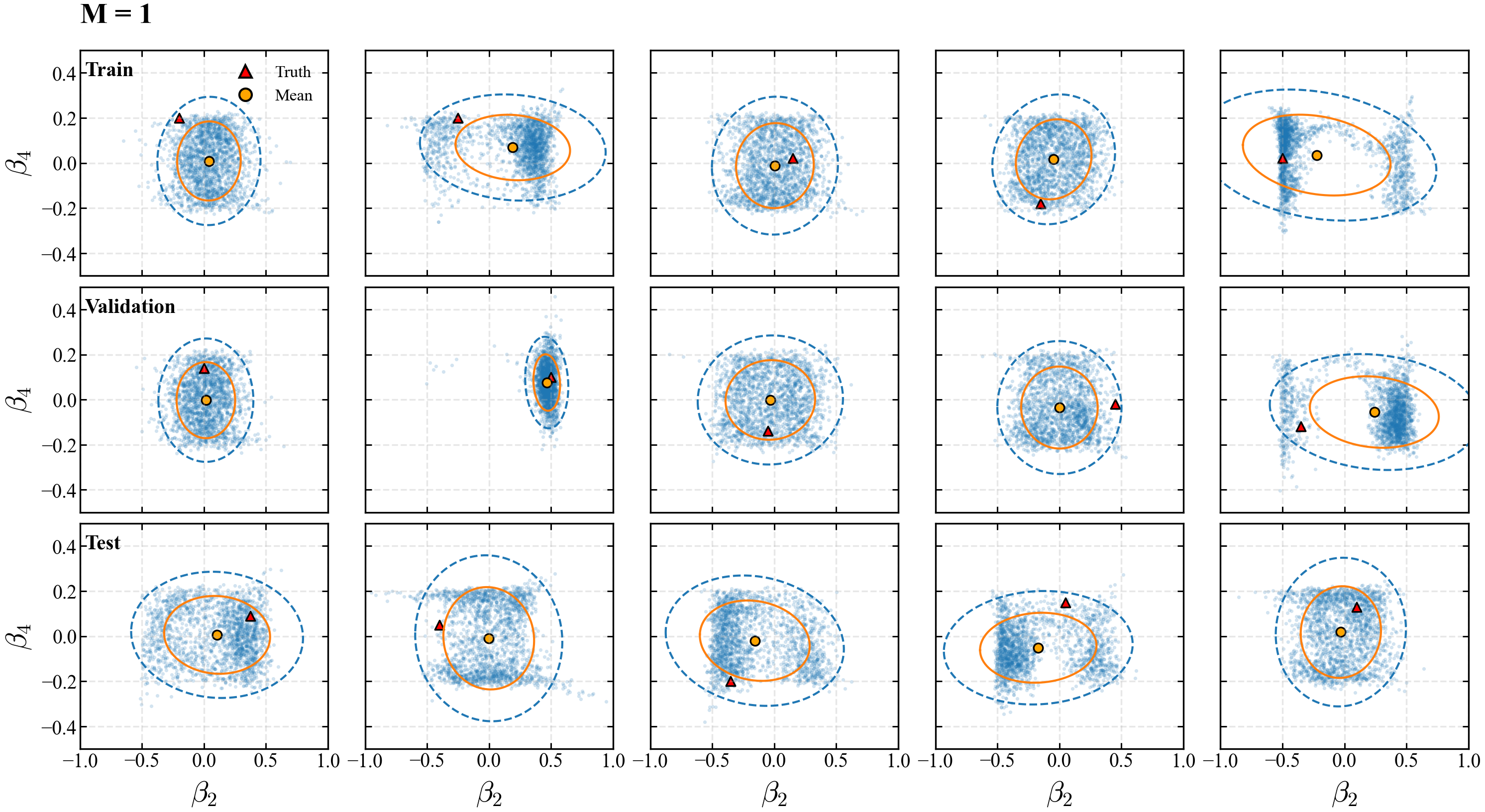}
		\caption{\label{SBI_Mahalanobis_M1}Two-dimensional posterior samples of $(\beta_2,\beta_4)$ obtained with SBI at $M=1$ (0--10\% centrality). The black cross denotes the true deformation parameters, the orange dot marks the posterior mean and the solid and dashed ellipses indicate the $1\sigma$ and $2\sigma$ Mahalanobis contours, respectively (0--10\% centrality).}
	\end{figure*}
	
	\begin{figure*}[htbp]
		\centering
		\includegraphics[scale=0.7]{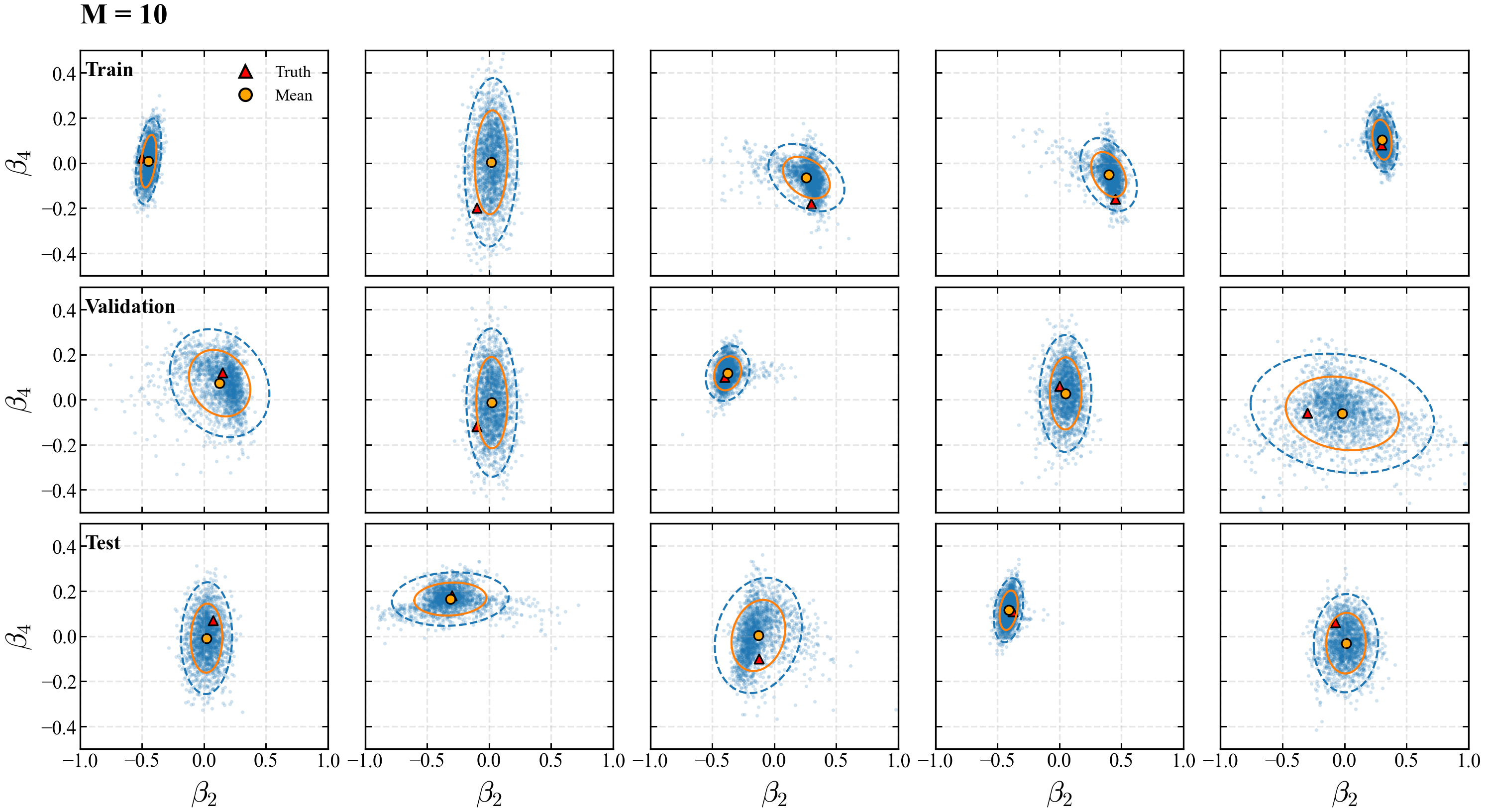}
		\caption{\label{SBI_Mahalanobis_M10}Same as Fig.~\ref{SBI_Mahalanobis_M1}, but for $M=10$ (0--10\% centrality).}
	\end{figure*}
	
	\begin{figure*}[htbp]
		\centering
		\includegraphics[scale=0.7]{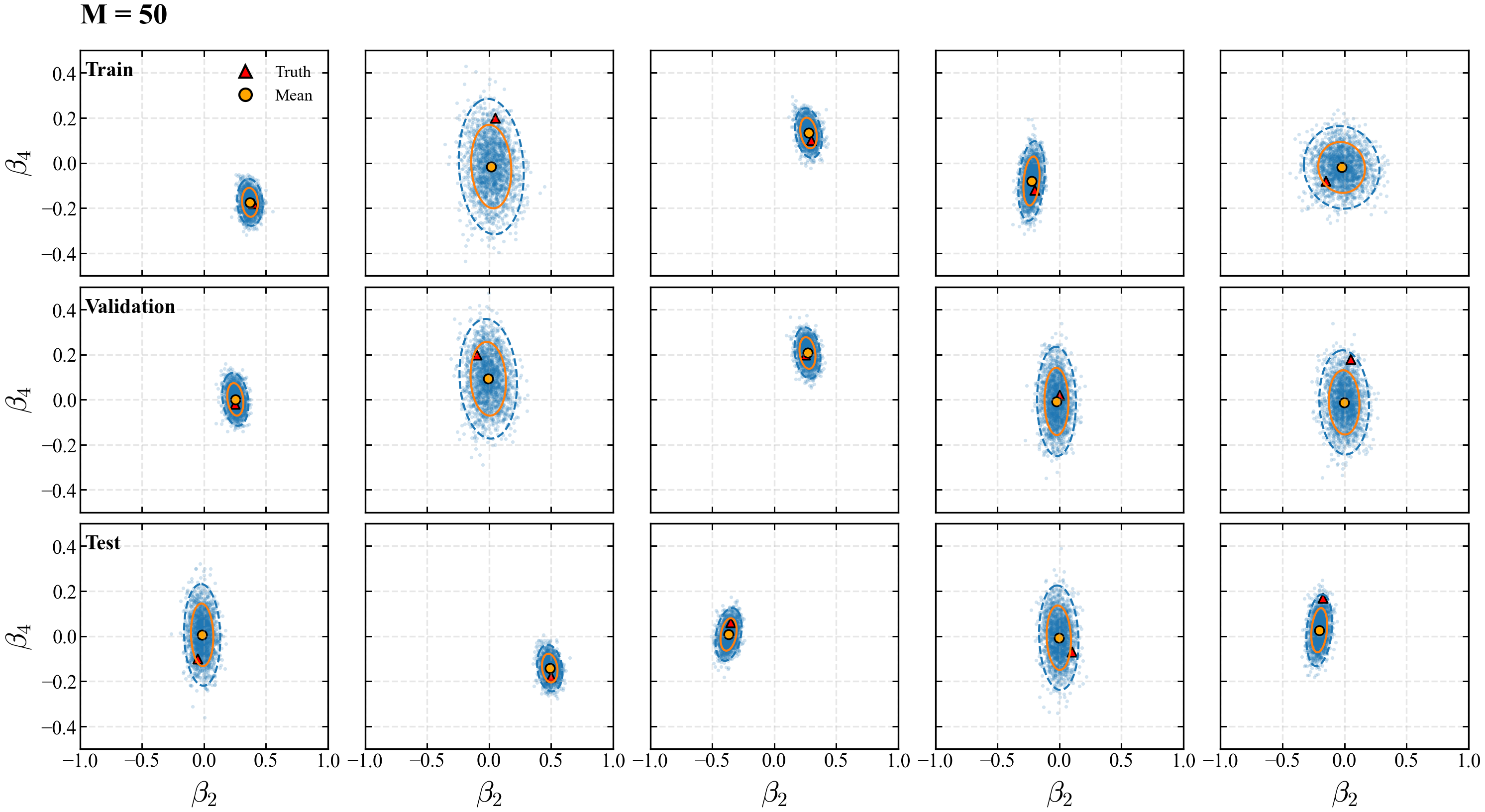}
		\caption{\label{SBI_Mahalanobis_M50}Same as Fig.~\ref{SBI_Mahalanobis_M1}, but for $M=50$ (0--10\% centrality).}
	\end{figure*}
	
	\begin{figure*}[htbp]
		\centering
		\includegraphics[scale=0.7]{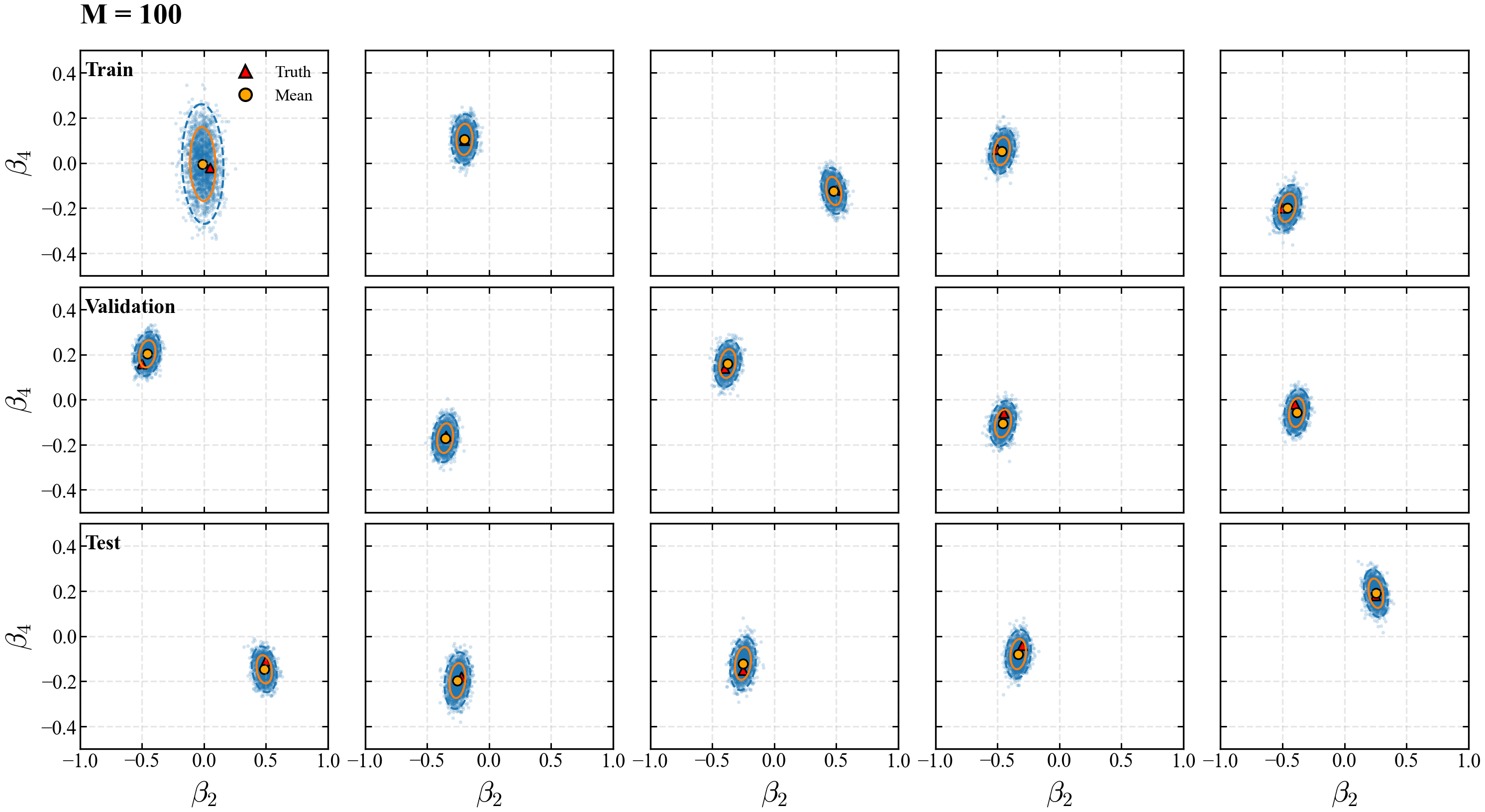}
		\caption{\label{SBI_Mahalanobis_M100}Same as Fig.~\ref{SBI_Mahalanobis_M1}, but for $M=100$ (0--10\% centrality).}
	\end{figure*}
	
	\begin{table*}[!htb]
		\renewcommand{\arraystretch}{1.5}
		\centering
		\caption{$R^2$ values for $\beta_2$ and $\beta_4$ between the SBI posterior means and the true deformation parameters for different bag sizes $M$ (0–-10\% centrality).}
		\begin{tabular}{>{\centering\arraybackslash}p{1.2cm}
				>{\centering\arraybackslash}p{2.5cm}
				>{\centering\arraybackslash}p{2.5cm}
				>{\centering\arraybackslash}p{2.5cm}
				>{\centering\arraybackslash}p{2.5cm}
				>{\centering\arraybackslash}p{2.5cm}
				>{\centering\arraybackslash}p{2.5cm}}
			\hline\hline
			\textbf{$M$} & 
			\textbf{Train $R^2$ ($\beta_2$)} & 
			\textbf{Train $R^2$ ($\beta_4$)} & 
			\textbf{Val $R^2$ ($\beta_2$)} & 
			\textbf{Val $R^2$ ($\beta_4$)} & 
			\textbf{Test $R^2$ ($\beta_2$)} &
			\textbf{Test $R^2$ ($\beta_4$)} \\
			\hline
			1   & 0.29372 & 0.11797 & 0.27184 & 0.10726 & 0.25802 & 0.10340 \\
			10  & 0.84790 & 0.44633 & 0.78571 & 0.42278 & 0.77898 & 0.40480 \\
			50  & 0.95818 & 0.66348 & 0.93062 & 0.62261 & 0.91787 & 0.60747 \\
			100 & 0.98095 & 0.75849 & 0.95216 & 0.67450 & 0.93926 & 0.65660 \\
			\hline\hline
		\end{tabular}
		\label{best_sbi_mean}
	\end{table*}
	
	\begin{figure*}[htbp]
		\centering
		\includegraphics[scale=0.45]{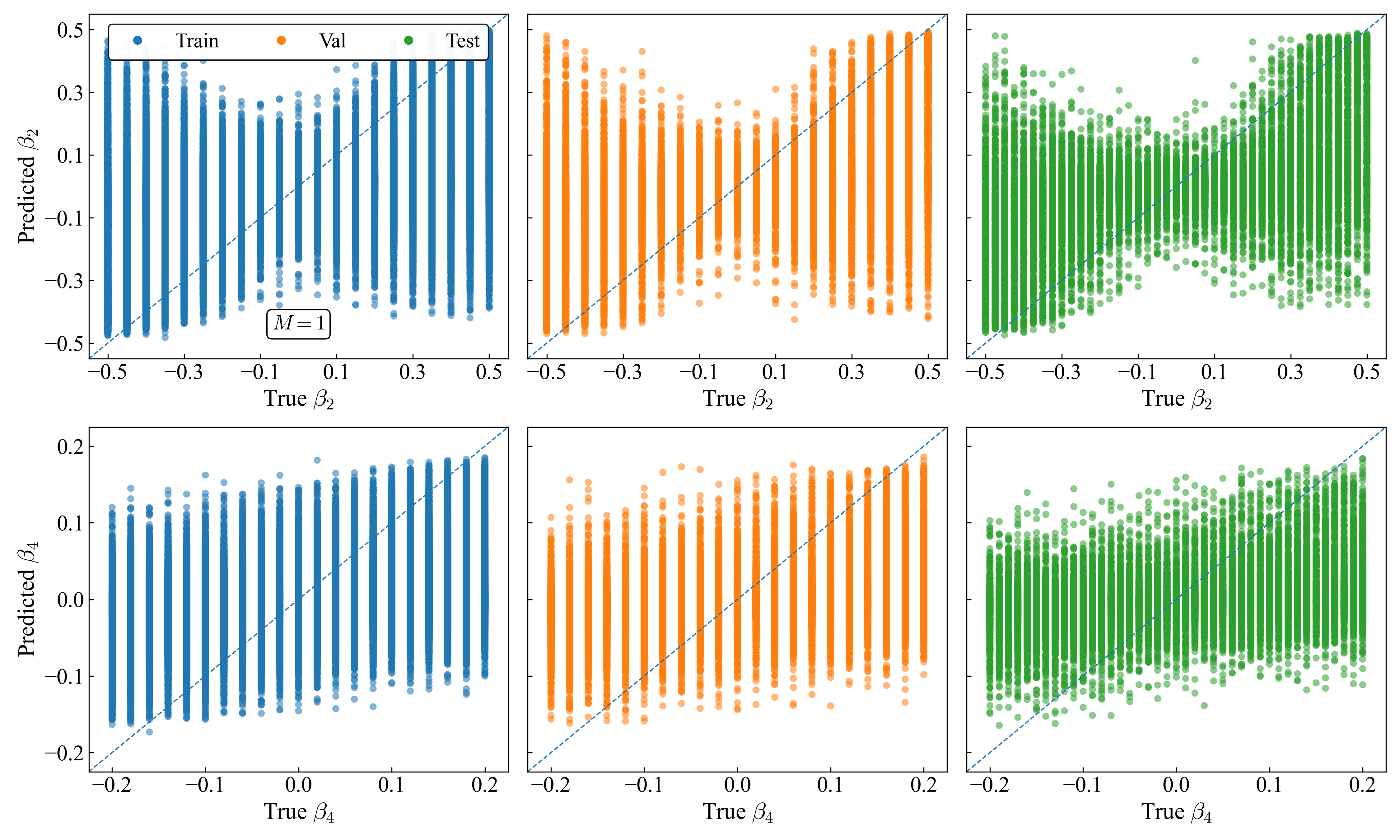}
		\caption{\label{sbi_mean_scatter_M1}SBI posterior means versus truth for $\beta_{2}$ (top) and $\beta_{4}$ (bottom) at $M=1$ (0--10\% centrality). The diagonal denotes perfect agreement.}
	\end{figure*}
	
	\begin{figure*}[htbp]
		\centering
		\includegraphics[scale=0.45]{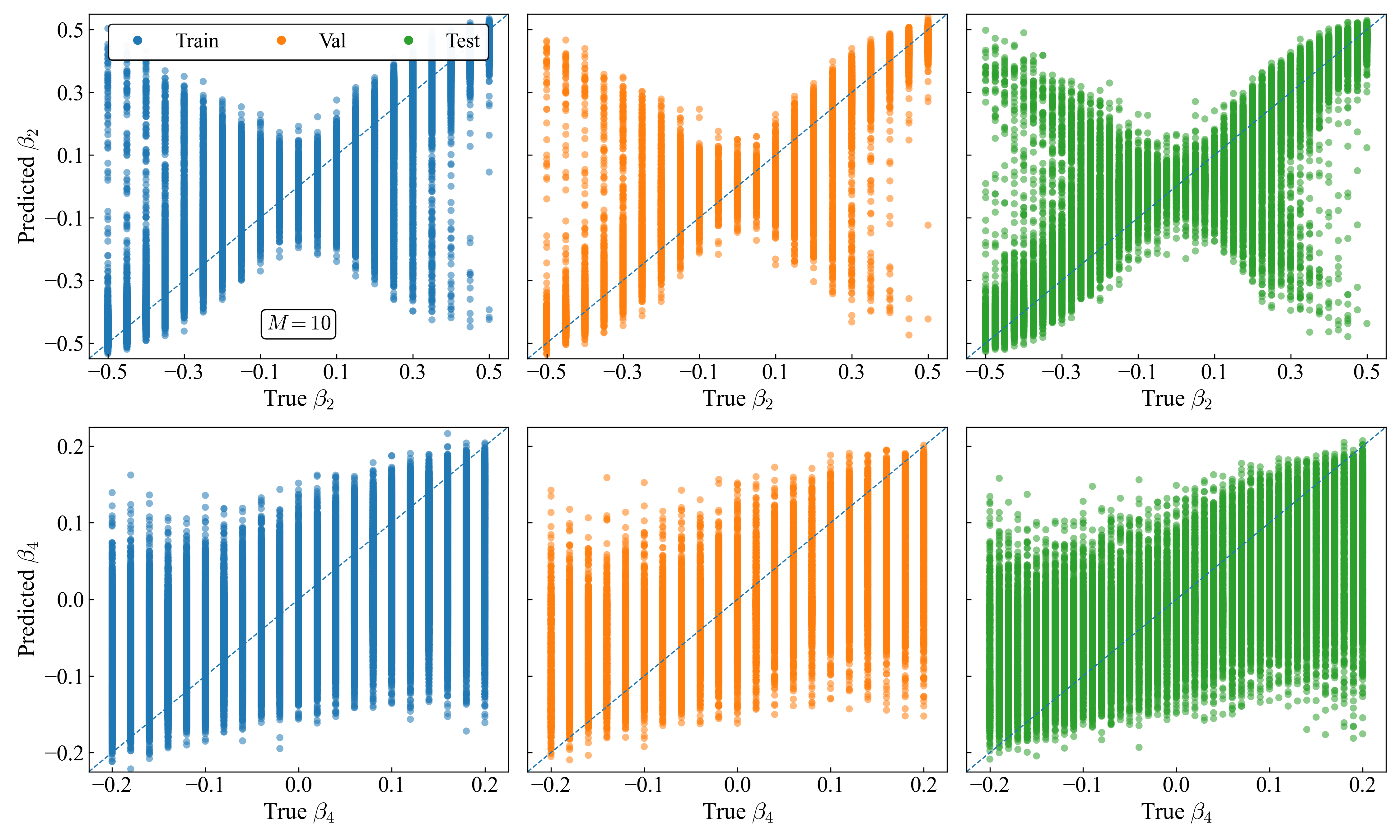}
		\caption{\label{sbi_mean_scatter_M10}Same as Fig.~\ref{sbi_mean_scatter_M1}, but for $M=10$ (0--10\% centrality).}
	\end{figure*}
	
	\begin{figure*}[htbp]
		\centering
		\includegraphics[scale=0.45]{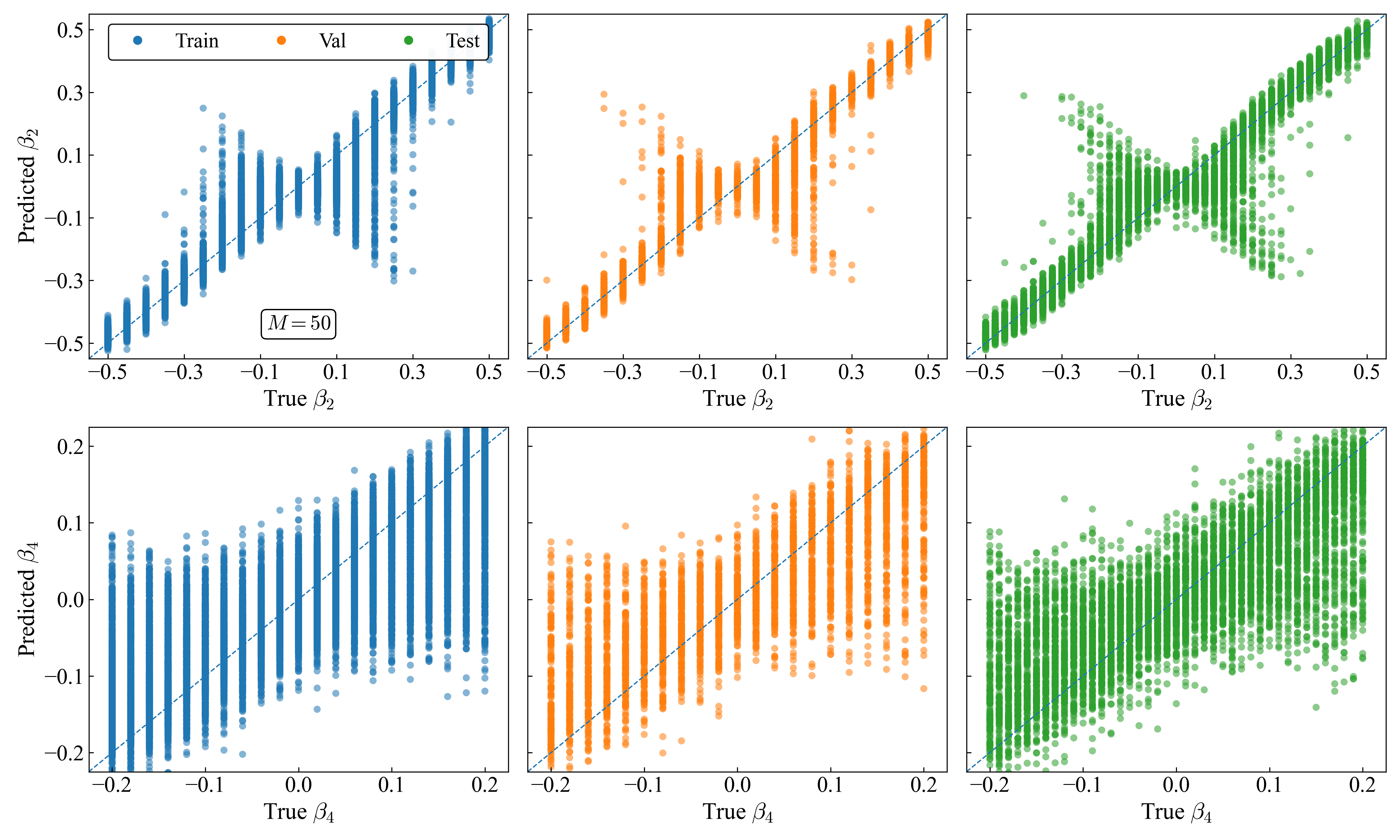}
		\caption{\label{sbi_mean_scatter_M50}Same as Fig.~\ref{sbi_mean_scatter_M1}, but for $M=50$ (0--10\% centrality).}
	\end{figure*}
	
	\begin{figure*}[htbp]
		\centering
		\includegraphics[scale=0.45]{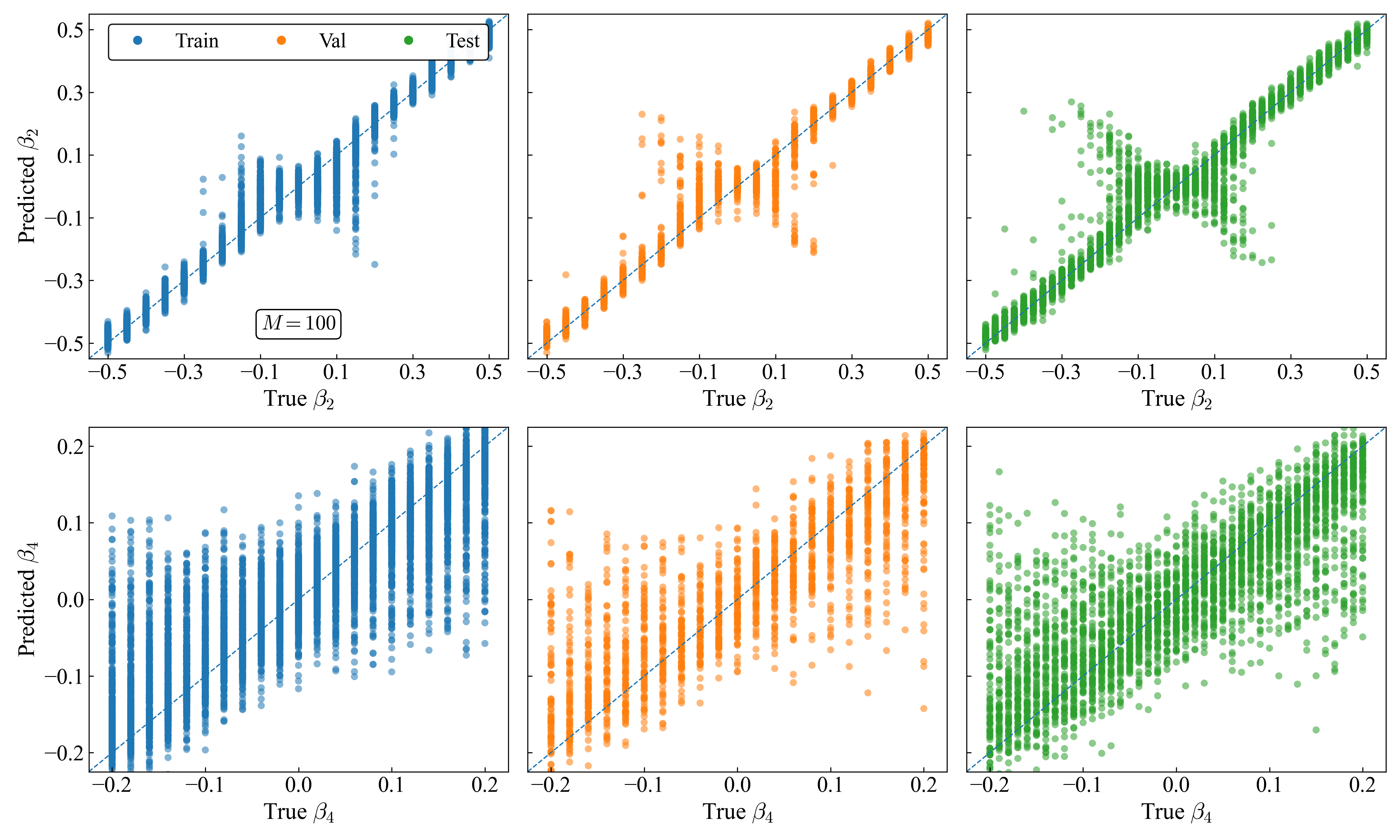}
		\caption{\label{sbi_mean_scatter_M100}Same as Fig.~\ref{sbi_mean_scatter_M1}, but for $M=100$ (0--10\% centrality).}
	\end{figure*}
	
	Figure~\ref{SBI_Train_Results} shows the training and validation negative log-likelihood (NLL) of the SBI model for different bag sizes $M = 1, 10, 50,$ and $100$ at 0--10\% centrality. Similar to the regression case, increasing $M$ leads to more stable optimization and a substantially lower NLL. For small bags, the validation curves fluctuate strongly and overfitting appears early, while larger bags yield smoother convergence and improved likelihood performance. The best overall behavior is obtained for $M = 100$, indicating that the conditional normalizing flow can more effectively learn the deformation-dependent posterior once event-by-event fluctuations in the entropy profiles are sufficiently suppressed.
	
	Table~\ref{best_sbi} summarizes the likelihood improvements for 0--10\% centrality achieved by the SBI model over a Gaussian baseline across different bag sizes $M$, revealing a clear monotonic trend: larger bags consistently yield higher likelihood gains for the training, validation and test sets. This behavior reflects the reduced influence of event-by-event fluctuations when more entropy-density profiles are aggregated, enabling the conditional flow to learn a sharper and more expressive representation of the deformation-dependent posterior. The quantitative increase from $M=1$ to $M=100$ highlights the strong dependence of posterior quality on the amount of statistical averaging provided to the model. Fig.~\ref{SBI_Cov} presents the empirical--nominal coverage curves of the inferred posteriors at different bag sizes $M$ \cite{Talts2020}. For $M=1$, all three data splits lie very close to the ideal diagonal, indicating that although the posterior is relatively diffuse, its uncertainty is well calibrated. Increasing the bag size to $M=10$ yields the best overall calibration: training, validation and test curves all track the ideal line with minimal deviation, suggesting that the deformation signal becomes sufficiently informative while the posterior uncertainty remains appropriately dispersed. For larger bags, $M=50$ and $M=100$, the coverage curves gradually move above the ideal line, especially for the training data, indicating mild statistical over-coverage. This suggests that, as $M$ increases and the deformation signal becomes sharper, the flow produces slightly overdispersed posteriors whose central credible regions are somewhat conservative, so that the nominal intervals contain the true values more often than intended. The probability integral transform (PIT) histograms in Fig.~\ref{SBI_PIT} provide complementary evidence \cite{Gneiting2007}. The PIT distribution for $M=1$ is nearly uniform, consistent with well-calibrated but relatively broad posteriors. For $M=10$, the histogram remains close to uniform with only mild right-skew, indicating that the posteriors are still well calibrated while the mean predictions become slightly sharper. For $M=50$ and $M=100$, the PIT histograms develop a pronounced hump-shaped structure---with an excess of central mass and deficits in the tails---indicating slightly overdispersed posteriors and conservative uncertainty quantification. The monotonic increase of the KS statistic from $M=1$ to $M=100$ quantitatively tracks these growing deviations from perfect uniformity, although its absolute values remain modest. Taken together, these results show that while $M=10$ achieves the most balanced posterior calibration, very large bag sizes sharpen the deformation signal to the point where the flow model becomes increasingly challenged to represent the full tail structure of the posterior, leading for $M\geq50$ to a combination of over-coverage and overdispersed PIT behavior, that is, more accurate mean predictions accompanied by slightly conservative uncertainty bands.
	
	Figures~\ref{SBI_Mahalanobis_M1}, \ref{SBI_Mahalanobis_M10}, \ref{SBI_Mahalanobis_M50}, and \ref{SBI_Mahalanobis_M100} show the two-dimensional posterior samples of $(\beta_{2},\beta_{4})$ for representative points in the parameter grid at 0--10\% centrality, together with their corresponding Mahalanobis ellipses at the $1\sigma$ and $2\sigma$ levels, defined by contours of constant quadratic form $(\mathbf{x}-\boldsymbol{\mu})^{\mathrm T}\Sigma^{-1}(\mathbf{x}-\boldsymbol{\mu})$ under a bivariate Gaussian approximation \cite{Mahalanobis1936,JohnsonWichern}. For $M=1$, the posteriors remain broad and visibly distorted, reflecting the strong event-by-event fluctuations present in single entropy-density profiles; the inferred means deviate non-negligibly from the true values and the posterior shapes vary substantially across different deformation points. Increasing the bag size to $M=10$ yields a marked improvement: the posteriors shrink significantly, become more elliptical, and center more closely around the true parameters, although noticeable variations in shape and orientation are still present for certain deformation configurations. At $M=50$, the posteriors become compact and highly regular, with consistent elliptical structure and means that closely track the targets across training, validation, and test sets; the true values typically lie well inside the $1\sigma$ ellipses, indicating slightly conservative credible regions in line with the mild over-coverage seen in the coverage and PIT diagnostics. For the largest bag size, $M=100$, the posteriors exhibit the most stable and concentrated behavior, with minimal scatter and strongly Gaussian shapes across all parameter points, while the true deformation parameters remain comfortably enclosed by the $1\sigma$ contours. These results clearly demonstrate the progressive refinement of posterior geometry as $M$ increases, highlighting the crucial role of multi-event aggregation in suppressing fluctuations and enabling accurate---though for $M=50$ and $M=100$ slightly conservative---reconstruction of $(\beta_{2},\beta_{4})$ through SBI.
	
	Table~\ref{best_sbi_mean} reports the $R^2$ values obtained from the posterior means of the SBI model. Compared with the regression results in Tab.~\ref{best_reg}, the posterior means exhibit equal or better accuracy across all bag sizes. At small $M$, both methods are limited by event-level fluctuations, but at moderate and large $M$ the SBI means achieve $R^2$ values that match or exceed those of direct regression, particularly for $\beta_4$ where the gains are most visible. For $M=100$, the SBI mean clearly outperforms the regression model on all data splits, demonstrating that once sufficient averaging is provided, the SBI posterior mean can surpass conventional regression in capturing the deformation trends. The results for other centrality classes in Appendix~\ref{appendixB} confirm the same trends. The corresponding scatter plots in Figs.~\ref{sbi_mean_scatter_M1}, \ref{sbi_mean_scatter_M10}, \ref{sbi_mean_scatter_M50} and \ref{sbi_mean_scatter_M100} show the same qualitative trends versus the regression scatter plots: increasing $M$ yields tighter alignment with the truth, $\beta_2$ remains easier to recover than $\beta_4$ and the $|\beta_2|$ degeneracy and improved accuracy near $\beta_2=0$ persist at small $M$.
	 
	In summary, the results from initial entropy-density profiles show the complementary behavior of the regression and SBI approaches, as well as their shared dependence on the bag size $M$ and collision centrality. Increasing $M$ systematically improves the performance of both methods, since averaging over more events suppresses stochastic fluctuations and enhances the effective deformation signal. At sufficiently large $M$, the point estimates produced by regression become highly accurate and the posterior means obtained from SBI match—or in some cases even slightly exceed—the regression performance, demonstrating that both approaches can reliably capture the central deformation trends once the deformation signal becomes sufficiently strong. However, regression remains fundamentally limited by its point-estimate nature and cannot represent the full uncertainty structure of $(\beta_2,\beta_4)$, whereas SBI provides full posterior distributions that enable uncertainty quantification and a more complete assessment of inference quality, even though accurately calibrating these posteriors becomes increasingly challenging as $M$ grows. Despite these improvements, both methods exhibit a marked centrality dependence: as shown in Appendix~\ref{appendixB}, inference quality systematically degrades toward more peripheral collisions, where fewer participant nucleons and weaker geometric imprint reduce the available deformation information. Overall, these findings demonstrate that large bag sizes are crucial for extracting deformation information and that while both methods face increasing difficulty at more peripheral collisions, the ability of SBI to recover full posteriors gives it a clear advantage over regression in robustness and uncertainty quantification.

	\section{Conclusion}\label{Sec4}
	In this work, we perform a comprehensive study of extracting nuclear deformation parameters $(\beta_2,\beta_4)$ from initial-state configurations in heavy-ion collisions using both regression and simulation-based inference (SBI) approaches. By first establishing a baseline with nucleon configurations sampled from a deformed Woods–Saxon distribution, we demonstrated that multi-event aggregation substantially enhances the identifiability of deformation parameters and is essential for suppressing stochastic fluctuations, thereby enabling strong predictive performance. Extending the analysis to entropy-density profiles generated by the T\raisebox{-0.5ex}{R}ENTo model, we observed a consistent trend: larger bag sizes $M$ lead to improved recovery of deformation signals across all centrality classes. While regression models effectively capture global deformation patterns, they are inherently limited by their point-estimate nature and lack of calibrated uncertainty. In contrast, SBI based on conditional normalizing flows provides full posterior distributions, enabling more comprehensive uncertainty quantification and a more reliable characterization of inference quality. Notably, for sufficiently large $M$, the posterior means from SBI achieve accuracy comparable to -- and in some cases exceeding -- that of regression, particularly for the more elusive hexadecapole parameter $\beta_4$.
	
	A consistent observation across all setups is the systematic deterioration of inference performance at more peripheral collisions. Peripheral collisions involve fewer participant nucleons and provide less geometric information about the bulk nuclear shape, leading to weaker deformation imprints and reduced identifiability. The representative results shown in Appendix~\ref{appendixB} confirm this degradation for both regression and SBI. Overall, our findings demonstrate that deformation information embedded in the initial-state geometry can be effectively amplified through multi-event averaging and is most robustly reconstructed using SBI. Future directions include extending this framework to hydrodynamic evolution, final-state observables and more realistic nuclear systems, thereby clarifying how deformation effects propagate through the full collision dynamics.
	
	\begin{acknowledgments}
		We thank Qian-Ru Lin for the computing resources. This work is supported by the National Natural Science Foundation of China (NSFC) under Grant Nos.~12535010 and 92570117, the Ministry of Science and Technology of China under Grant No.~2024YFA1611004, the Shenzhen Peacock Fund under Grant No.~2023TC0179, the CUHK-Shenzhen University Development Fund under Grant Nos.~UDF01003041 and UDF03003041, the China Scholarship Council (CSC) under Grant No.~202406770003, and the China Postdoctoral Science Foundation under Grant No.~2025M781519.
	\end{acknowledgments}

	\appendix
	\section{Entropy-profile comparison}\label{appendixA}
	\begin{figure*}[htbp]
		\centering
		\includegraphics[scale=0.8]{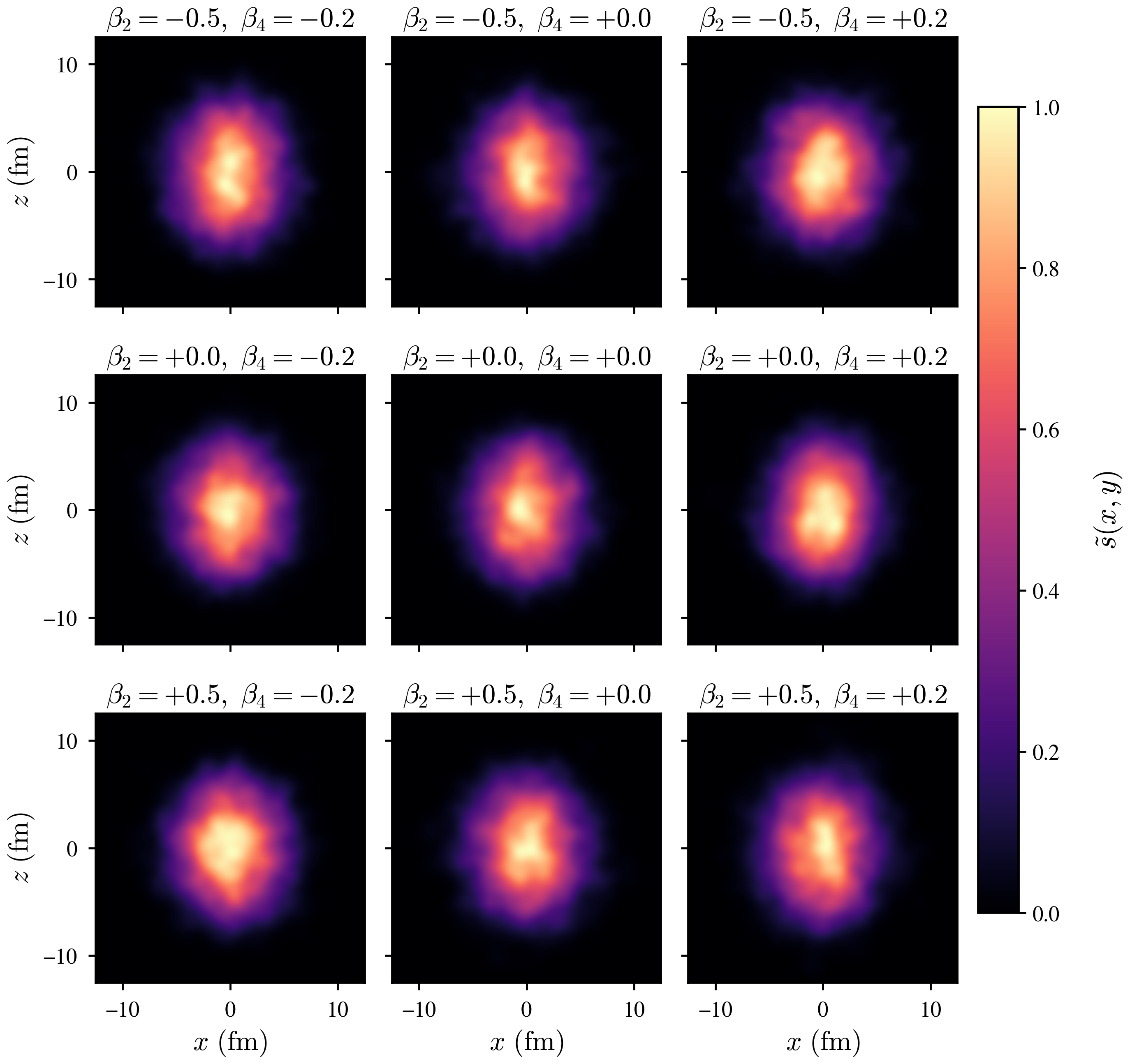}
		\caption{\label{EP}Entropy-density profiles for representative $(\beta_2,\beta_4)$ values, summed over 100 events (0--100\%).}
	\end{figure*}
	
	Figure~\ref{EP} shows representative entropy-density profiles for different $(\beta_2,\beta_4)$ values, summed over 100 events at 0--100\% centrality. The profiles remain visually very similar across different deformation parameters, indicating that direct visual inspection is insufficient for reliable discrimination.
	
	\section{Results for the remaining centrality bins}\label{appendixB}
	
	Tables~\ref{best_reg_30_40}--\ref{best_sbi_mean_0_100} collect the regression and SBI results for the additional centrality bins 30--40\%, 60--70\%, 90--100\% and 0--100\%, which are omitted from the main text for brevity. These tables follow the same format as those shown for the 0--10\% bin and illustrate the expected trend that deformation inference becomes progressively more difficult toward higher centralities due to the reduced participant count and weaker geometric imprint.

	\begin{table*}[!htb]
		\renewcommand{\arraystretch}{1.5}
		\centering
		\caption{Same as Tab.~\ref{best_reg}, but for 30--40\%.}
		\begin{tabular}{>{\centering\arraybackslash}p{1.2cm}
				>{\centering\arraybackslash}p{2.5cm}
				>{\centering\arraybackslash}p{2.5cm}
				>{\centering\arraybackslash}p{2.5cm}
				>{\centering\arraybackslash}p{2.5cm}
				>{\centering\arraybackslash}p{2.5cm}
				>{\centering\arraybackslash}p{2.5cm}}
			\hline\hline
			\textbf{$M$} & 
			\textbf{Train $R^2$ ($\beta_2$)} & 
			\textbf{Train $R^2$ ($\beta_4$)} & 
			\textbf{Val $R^2$ ($\beta_2$)} & 
			\textbf{Val $R^2$ ($\beta_4$)} & 
			\textbf{Test $R^2$ ($\beta_2$)} &
			\textbf{Test $R^2$ ($\beta_4$)} \\
			\hline
			1   & 0.16362 & 0.04844 & 0.15182 & 0.04521 & 0.14016 & 0.04528 \\
			10  & 0.69425 & 0.23845 & 0.58658 & 0.20633 & 0.56980 & 0.20323 \\
			50  & 0.91346 & 0.46296 & 0.77207 & 0.37950 & 0.74967 & 0.35716 \\
			100 & 0.94561 & 0.56205 & 0.83105 & 0.41367 & 0.78257 & 0.38620 \\
			\hline\hline
		\end{tabular}
		\label{best_reg_30_40}
	\end{table*}
	
	\begin{table*}[!htb]
		\renewcommand{\arraystretch}{1.5}
		\centering
		\caption{Same as Tab.~\ref{best_sbi}, but for 30--40\%.}
		\begin{tabular}{>{\centering\arraybackslash}p{2.4cm}
				>{\centering\arraybackslash}p{5cm}
				>{\centering\arraybackslash}p{5cm}
				>{\centering\arraybackslash}p{5cm}}
			\hline\hline
			\textbf{$M$} & 
			\textbf{Train LL Gain (vs. Baseline)} & 
			\textbf{Val LL Gain (vs. Baseline)} & 
			\textbf{Test LL Gain (vs. Baseline)} \\
			\hline
			1   & 0.66331 & 0.62579 & 0.59620  \\
			10  & 1.75998 & 1.52469 & 1.49647  \\
			50  & 2.28308 & 1.83118 & 1.73420  \\
			100 & 2.38508 & 2.03145 & 1.95800  \\
			\hline\hline
		\end{tabular}
		\label{best_sbi_30_40}
	\end{table*}
	
	\begin{table*}[!htb]
		\renewcommand{\arraystretch}{1.5}
		\centering
		\caption{Same as Tab.~\ref{best_sbi_mean}, but for 30--40\%.}
		\begin{tabular}{>{\centering\arraybackslash}p{1.2cm}
				>{\centering\arraybackslash}p{2.5cm}
				>{\centering\arraybackslash}p{2.5cm}
				>{\centering\arraybackslash}p{2.5cm}
				>{\centering\arraybackslash}p{2.5cm}
				>{\centering\arraybackslash}p{2.5cm}
				>{\centering\arraybackslash}p{2.5cm}}
			\hline\hline
			\textbf{$M$} & 
			\textbf{Train $R^2$ ($\beta_2$)} & 
			\textbf{Train $R^2$ ($\beta_4$)} & 
			\textbf{Val $R^2$ ($\beta_2$)} & 
			\textbf{Val $R^2$ ($\beta_4$)} & 
			\textbf{Test $R^2$ ($\beta_2$)} &
			\textbf{Test $R^2$ ($\beta_4$)} \\
			\hline
			1   & 0.19107 & 0.07251 & 0.17460 & 0.06166 & 0.16487 & 0.05684 \\
			10  & 0.70142 & 0.32506 & 0.60984 & 0.25694 & 0.59394 & 0.25018 \\
			50  & 0.92050 & 0.53751 & 0.79934 & 0.43271 & 0.78692 & 0.41326 \\
			100 & 0.92258 & 0.55798 & 0.84776 & 0.46265 & 0.81630 & 0.45093 \\
			\hline\hline
		\end{tabular}
		\label{best_sbi_mean_30_40}
	\end{table*}
	
	\begin{table*}[!htb]
		\renewcommand{\arraystretch}{1.5}
		\centering
		\caption{Same as Tab.~\ref{best_reg}, but for 60--70\%.}
		\begin{tabular}{>{\centering\arraybackslash}p{1.2cm}
				>{\centering\arraybackslash}p{2.5cm}
				>{\centering\arraybackslash}p{2.5cm}
				>{\centering\arraybackslash}p{2.5cm}
				>{\centering\arraybackslash}p{2.5cm}
				>{\centering\arraybackslash}p{2.5cm}
				>{\centering\arraybackslash}p{2.5cm}}
			\hline\hline
			\textbf{$M$} & 
			\textbf{Train $R^2$ ($\beta_2$)} & 
			\textbf{Train $R^2$ ($\beta_4$)} & 
			\textbf{Val $R^2$ ($\beta_2$)} & 
			\textbf{Val $R^2$ ($\beta_4$)} & 
			\textbf{Test $R^2$ ($\beta_2$)} &
			\textbf{Test $R^2$ ($\beta_4$)} \\
			\hline
			1   & 0.07664 & 0.02504 & 0.06017 & 0.02534 & 0.06116 & 0.01922 \\
			10  & 0.39620 & 0.10103 & 0.30101 & 0.09717 & 0.29180 & 0.09155 \\
			50  & 0.66292 & 0.18714 & 0.49488 & 0.17715 & 0.48339 & 0.15387 \\
			100 & 0.73615 & 0.26239 & 0.55405 & 0.21366 & 0.52556 & 0.20863 \\
			\hline\hline
		\end{tabular}
		\label{best_reg_60-70}
	\end{table*}
	
	\begin{table*}[!htb]
		\renewcommand{\arraystretch}{1.5}
		\centering
		\caption{Same as Tab.~\ref{best_sbi}, but for 60--70\%.}
		\begin{tabular}{>{\centering\arraybackslash}p{2.4cm}
				>{\centering\arraybackslash}p{5cm}
				>{\centering\arraybackslash}p{5cm}
				>{\centering\arraybackslash}p{5cm}}
			\hline\hline
			\textbf{$M$} & 
			\textbf{Train LL Gain (vs. Baseline)} & 
			\textbf{Val LL Gain (vs. Baseline)} & 
			\textbf{Test LL Gain (vs. Baseline)} \\
			\hline
			1   & 0.49869 & 0.46321 & 0.43682  \\
			10  & 1.20598 & 1.12463 & 1.08396  \\
			50  & 2.10833 & 1.76098 & 1.70491  \\
			100 & 2.29309 & 1.92498 & 1.81570  \\
			\hline\hline
		\end{tabular}
		\label{best_sbi_60-70}
	\end{table*}
	
	\begin{table*}[!htb]
		\renewcommand{\arraystretch}{1.5}
		\centering
		\caption{Same as Tab.~\ref{best_sbi_mean}, but for 60--70\%.}
		\begin{tabular}{>{\centering\arraybackslash}p{1.2cm}
				>{\centering\arraybackslash}p{2.5cm}
				>{\centering\arraybackslash}p{2.5cm}
				>{\centering\arraybackslash}p{2.5cm}
				>{\centering\arraybackslash}p{2.5cm}
				>{\centering\arraybackslash}p{2.5cm}
				>{\centering\arraybackslash}p{2.5cm}}
			\hline\hline
			\textbf{$M$} & 
			\textbf{Train $R^2$ ($\beta_2$)} & 
			\textbf{Train $R^2$ ($\beta_4$)} & 
			\textbf{Val $R^2$ ($\beta_2$)} & 
			\textbf{Val $R^2$ ($\beta_4$)} & 
			\textbf{Test $R^2$ ($\beta_2$)} &
			\textbf{Test $R^2$ ($\beta_4$)} \\
			\hline
			1   & 0.08102 & 0.03664 & 0.06624 & 0.02799 & 0.06452 & 0.02155 \\
			10  & 0.39252 & 0.14218 & 0.36769 & 0.12698 & 0.34945 & 0.11412 \\
			50  & 0.69608 & 0.31933 & 0.55242 & 0.24794 & 0.58002 & 0.22115 \\
			100 & 0.77546 & 0.37326 & 0.62196 & 0.21944 & 0.64044 & 0.20245 \\
			\hline\hline
		\end{tabular}
		\label{best_sbi_mean_60_70}
	\end{table*}
	
	\begin{table*}[!htb]
		\renewcommand{\arraystretch}{1.5}
		\centering
		\caption{Same as Tab.~\ref{best_reg}, but for 90--100\%.}
		\begin{tabular}{>{\centering\arraybackslash}p{1.2cm}
				>{\centering\arraybackslash}p{2.5cm}
				>{\centering\arraybackslash}p{2.5cm}
				>{\centering\arraybackslash}p{2.5cm}
				>{\centering\arraybackslash}p{2.5cm}
				>{\centering\arraybackslash}p{2.5cm}
				>{\centering\arraybackslash}p{2.5cm}}
			\hline\hline
			\textbf{$M$} & 
			\textbf{Train $R^2$ ($\beta_2$)} & 
			\textbf{Train $R^2$ ($\beta_4$)} & 
			\textbf{Val $R^2$ ($\beta_2$)} & 
			\textbf{Val $R^2$ ($\beta_4$)} & 
			\textbf{Test $R^2$ ($\beta_2$)} &
			\textbf{Test $R^2$ ($\beta_4$)} \\
			\hline
			1   & 0.00577 & 0.00323 & 0.00420 & 0.00382 & 0.00339 & 0.00443 \\
			10  & 0.05444 & 0.02725 & 0.04521 & 0.02211 & 0.03554 & 0.02376 \\
			50  & 0.15432 & 0.03902 & 0.09888 & 0.03088 & 0.08900 & 0.03289 \\
			100 & 0.29098 & 0.06941 & 0.11695 & 0.04534 & 0.07419 & 0.04508 \\
			\hline\hline
		\end{tabular}
		\label{best_reg_90-100}
	\end{table*}
	
	\begin{table*}[!htb]
		\renewcommand{\arraystretch}{1.5}
		\centering
		\caption{Same as Tab.~\ref{best_sbi}, but for 90--100\%.}
		\begin{tabular}{>{\centering\arraybackslash}p{2.4cm}
				>{\centering\arraybackslash}p{5cm}
				>{\centering\arraybackslash}p{5cm}
				>{\centering\arraybackslash}p{5cm}}
			\hline\hline
			\textbf{$M$} & 
			\textbf{Train LL Gain (vs. Baseline)} & 
			\textbf{Val LL Gain (vs. Baseline)} & 
			\textbf{Test LL Gain (vs. Baseline)} \\
			\hline
			1   & 0.41404 & 0.36513 & 0.36101  \\
			10  & 0.51972 & 0.48891 & 0.48324  \\
			50  & 0.98936 & 0.88769 & 0.88438  \\
			100 & 1.31604 & 1.08473 & 1.04499  \\
			\hline\hline
		\end{tabular}
		\label{best_sbi_90-100}
	\end{table*}
	
	\begin{table*}[!htb]
		\renewcommand{\arraystretch}{1.5}
		\centering
		\caption{Same as Tab.~\ref{best_sbi_mean}, but for 90--100\%.}
		\begin{tabular}{>{\centering\arraybackslash}p{1.2cm}
				>{\centering\arraybackslash}p{2.5cm}
				>{\centering\arraybackslash}p{2.5cm}
				>{\centering\arraybackslash}p{2.5cm}
				>{\centering\arraybackslash}p{2.5cm}
				>{\centering\arraybackslash}p{2.5cm}
				>{\centering\arraybackslash}p{2.5cm}}
			\hline\hline
			\textbf{$M$} & 
			\textbf{Train $R^2$ ($\beta_2$)} & 
			\textbf{Train $R^2$ ($\beta_4$)} & 
			\textbf{Val $R^2$ ($\beta_2$)} & 
			\textbf{Val $R^2$ ($\beta_4$)} & 
			\textbf{Test $R^2$ ($\beta_2$)} &
			\textbf{Test $R^2$ ($\beta_4$)} \\
			\hline
			1   & 0.01629 & 0.00906 & 0.00043 & 0.00159 & 0.00123 & 0.00016 \\
			10  & 0.06233 & 0.03691 & 0.05597 & 0.02462 & 0.04384 & 0.02680 \\
			50  & 0.20316 & 0.10573 & 0.16144 & 0.07602 & 0.15584 & 0.08015 \\
			100 & 0.26876 & 0.14258 & 0.21602 & 0.10158 & 0.20052 & 0.11154 \\
			\hline\hline
		\end{tabular}
		\label{best_sbi_mean_90_100}
	\end{table*}
	
	\begin{table*}[!htb]
		\renewcommand{\arraystretch}{1.5}
		\centering
		\caption{Same as Tab.~\ref{best_reg}, but for 0--100\%.}
		\begin{tabular}{>{\centering\arraybackslash}p{1.2cm}
				>{\centering\arraybackslash}p{2.5cm}
				>{\centering\arraybackslash}p{2.5cm}
				>{\centering\arraybackslash}p{2.5cm}
				>{\centering\arraybackslash}p{2.5cm}
				>{\centering\arraybackslash}p{2.5cm}
				>{\centering\arraybackslash}p{2.5cm}}
			\hline\hline
			\textbf{$M$} & 
			\textbf{Train $R^2$ ($\beta_2$)} & 
			\textbf{Train $R^2$ ($\beta_4$)} & 
			\textbf{Val $R^2$ ($\beta_2$)} & 
			\textbf{Val $R^2$ ($\beta_4$)} & 
			\textbf{Test $R^2$ ($\beta_2$)} &
			\textbf{Test $R^2$ ($\beta_4$)} \\
			\hline
			1   & 0.11552 & 0.02808 & 0.10528 & 0.02535 & 0.09157 & 0.02064 \\
			10  & 0.61891 & 0.15695 & 0.47809 & 0.13213 & 0.45499 & 0.12458 \\
			50  & 0.91568 & 0.42134 & 0.76072 & 0.31486 & 0.74631 & 0.29948 \\
			100 & 0.94555 & 0.47475 & 0.83475 & 0.38928 & 0.81726 & 0.36808 \\
			\hline\hline
		\end{tabular}
		\label{best_reg_0-100}
	\end{table*}
	
	\begin{table*}[!htb]
		\renewcommand{\arraystretch}{1.5}
		\centering
		\caption{Same as Tab.~\ref{best_sbi}, but for 0--100\%.}
		\begin{tabular}{>{\centering\arraybackslash}p{2.4cm}
				>{\centering\arraybackslash}p{5cm}
				>{\centering\arraybackslash}p{5cm}
				>{\centering\arraybackslash}p{5cm}}
			\hline\hline
			\textbf{$M$} & 
			\textbf{Train LL Gain (vs. Baseline)} & 
			\textbf{Val LL Gain (vs. Baseline)} & 
			\textbf{Test LL Gain (vs. Baseline)} \\
			\hline
			1   & 0.59296 & 0.52724 & 0.50357  \\
			10  & 1.53668 & 1.33133 & 1.28613  \\
			50  & 1.98768 & 1.68208 & 1.56881  \\
			100 & 2.35923 & 1.82520 & 1.61382 \\
			\hline\hline
		\end{tabular}
		\label{best_sbi_0-100}
	\end{table*}
	
	\begin{table*}[!htb]
		\renewcommand{\arraystretch}{1.5}
		\centering
		\caption{Same as Tab.~\ref{best_sbi_mean}, but for 0--100\%.}
		\begin{tabular}{>{\centering\arraybackslash}p{1.2cm}
				>{\centering\arraybackslash}p{2.5cm}
				>{\centering\arraybackslash}p{2.5cm}
				>{\centering\arraybackslash}p{2.5cm}
				>{\centering\arraybackslash}p{2.5cm}
				>{\centering\arraybackslash}p{2.5cm}
				>{\centering\arraybackslash}p{2.5cm}}
			\hline\hline
			\textbf{$M$} & 
			\textbf{Train $R^2$ ($\beta_2$)} & 
			\textbf{Train $R^2$ ($\beta_4$)} & 
			\textbf{Val $R^2$ ($\beta_2$)} & 
			\textbf{Val $R^2$ ($\beta_4$)} & 
			\textbf{Test $R^2$ ($\beta_2$)} &
			\textbf{Test $R^2$ ($\beta_4$)} \\
			\hline
			1   & 0.15151 & 0.06060 & 0.12862 & 0.04047 & 0.11285 & 0.03607 \\
			10  & 0.61603 & 0.28575 & 0.52013 & 0.21728 & 0.50525 & 0.20549 \\
			50  & 0.86133 & 0.46258 & 0.77472 & 0.38556 & 0.75571 & 0.37014 \\
			100 & 0.92928 & 0.55941 & 0.82586 & 0.45922 & 0.80635 & 0.43546 \\
			\hline\hline
		\end{tabular}
		\label{best_sbi_mean_0_100}
	\end{table*}
	
	\bibliography{Ref_v3.bib}
	
\end{document}